\documentclass[aps,pre,twocolumn,floatfix,10pt,showpacs,longbibliography]{revtex4-1}

\usepackage{graphicx}
\usepackage{verbatim}
\usepackage{comment}
\usepackage{amssymb}
\usepackage{enumerate}
\usepackage{psfrag}
\usepackage{pstricks}
\usepackage{color}
\usepackage{epstopdf}
\usepackage{MnSymbol}
\usepackage{ wasysym }
\usepackage{float}

\usepackage[colorlinks=true,urlcolor=blue,linkcolor=blue,citecolor=blue]{hyperref}

\usepackage{soul}

\emergencystretch=\maxdimen
\begin{document}

\title{Unsupervised machine learning account of magnetic transitions in the Hubbard model}

\author{Kelvin Ch'ng}
\affiliation{Department of Physics and Astronomy, San Jos\'{e} State University, San Jos\'{e}, CA 95192, USA}
\author{Nick Vazquez}
\affiliation{Department of Physics and Astronomy, San Jos\'{e} State University, San Jos\'{e}, CA 95192, USA}
\author{Ehsan Khatami}
\affiliation{Department of Physics and Astronomy, San Jos\'{e} State University, San Jos\'{e}, CA 95192, USA}

\begin{abstract}
We employ several unsupervised machine learning techniques, including autoencoders, random trees embedding, and 
t-distributed stochastic neighboring ensemble (t-SNE), to reduce the dimensionality of, and therefore classify, 
raw (auxiliary) spin configurations generated, through Monte Carlo simulations of small clusters, for the Ising and 
Fermi-Hubbard models at finite temperatures. Results from a convolutional autoencoder for the three-dimensional Ising 
model can be shown to produce the magnetization and the susceptibility as a function of temperature with a high degree of accuracy. 
Quantum fluctuations distort this picture and prevent us from making such connections between the output of the autoencoder and 
physical observables for the Hubbard model. However, we are able to define an indicator based on the output 
of the t-SNE algorithm that shows a near perfect agreement with the antiferromagnetic structure factor of the model in 
two and three spatial dimensions in the weak-coupling regime. t-SNE also predicts a transition to the canted antiferromagnetic 
phase for the three-dimensional model when a strong magnetic field is present. We show that these techniques cannot be
expected to work away from half filling when the ``sign problem" in quantum Monte Carlo simulations is present.
\end{abstract}

\maketitle

\section{Introduction}

Machine learning has emerged as an unconventional tool to gain insight into properties of many-body physics. In 
2014, Louis-Fran\c{c}ois et al.,~\cite{l_arsenault_14} used support vector machines, a type of supervised learning models, 
to obtain Green's function of the Anderson impurity model. Supervised machine learning 
techniques based on artificial neural networks were used later in a groundbreaking work to classify phases of models in statistical mechanics
and condensed matter physics~\cite{j_carrasquilla_16}. Shortly after, other groups expanded the application of these techniques 
to identify phase transitions, including to topological phases, in quantum many-body systems at zero or finite temperatures~
\cite{e_vanNieuwenburg_17,p_broecker_16,k_chng_17,y_zhang_17a,y_zhang_17b}. 

Parallel efforts demonstrated the power of restricted Boltzmann machines, simple artificial neural networks with one visible layer 
corresponding to the physical system and one hidden layer, in learning thermodynamics of Ising models~\cite{Torlai2016}, producing starting 
points for variational quantum Monte Carlo that are superior to those from 
conventional methods~\cite{Carleo2016}, performing tomography for many-body 
quantum states~\cite{g_torlai_17}, and constructing topological states~\cite{d_deng_16}. Interesting connections between artificial 
neural networks and more conventional methods in condensed matter physics have also been uncovered~
\cite{p_mehta_14,m_stoudenmire_16}

Unsupervised machine learning techniques, on the other hand, have so far been mostly used to classify 
phases of classical model in many-body physics. For example, 
t-distributed stochastic neighboring ensemble (t-SNE) technique~\cite{tSNE_Paper,tSNE_Link,tSNE_Code} 
was used in Ref.~\cite{j_carrasquilla_16} to cluster spin configurations and visualize the 
phase transition of the two-dimensional (2D) Ising model. Later, Lei Wang~\cite{Wang2016} applied principle component analysis 
(PCA)~\cite{pca} to 
obtain low-dimensional representations of Ising spin configurations and make connections between principle components and physical 
observables such as the magnetization and the susceptibility, conventionally used to determine critical phenomena. His work was recently 
followed up by other groups who provided a more detailed examination of the PCA and other techniques applied to various classical 
models, including those on frustrated geometries~\cite{s_wetzel_17,w_hu_17,c_wang_17}. PCA has also been 
applied to quantum systems~\cite{e_vanNieuwenburg_17},
however, the 2D visualization of the spin configuration for the random-field Heisenberg model did not 
produce any useful features.  A proposal for a different type of unsupervised machine learning for quantum many-body systems, which 
combines two-point function calculations with convolutional neural networks, is also recently put forth~\cite{p_broecker_17}. 

Here we employ several nonlinear unsupervised machine learning methods, including fully-connected and convolutional 
autoencoders~\cite{h_bourlard_88,g_hinton_06,f_chollet_16}, 
random trees embedding~\cite{p_geurts_06,f_moosmann_07,randomtrees}, 
and the t-SNE, to reduce the dimensionality of raw auxiliary spin (also known as auxiliary field) configurations 
generated during quantum Monte Carlo simulations of the two- and three-dimensional (3D) Fermi-Hubbard models~\cite{j_hubbard_63}. 
We focus on the finite-temperature 
magnetic phase transitions of the model in three or the corresponding crossover in two spatial dimensions. Therefore, 
most of the data are generated at half filling, 
where there are on average one fermion per lattice site. We visualize the outcomes and 
look for features in the dimensionally-reduced configurations that may correlate with 
physical observables or signal phase transitions/crossovers. We work with unlabeled data during 
learning; we do not use any knowledge of the model parameters or temperature each configuration 
corresponds to in our analysis, nor do we use any information about the location of the phase transitions or crossovers
during learning.

We start, however, with the classical Ising model on a 3D cubic lattice and benchmark the outcome of our convolutional 
autoencoder using spin configurations 
generated in a Monte Carlo simulation in a range of temperatures on both sides of the phase transition. 
We are able to define indicators that closely resemble the magnetization or the susceptibility. 
We then generalize the neural network to accommodate for the additional imaginary time axis in the auxiliary 
spin configurations of the Hubbard models and show that quantum fluctuations as well as the O(3) symmetry of the model in this case lead 
to low-dimensional visualizations that are fuzzier than their classical counterparts, 
although useful indicators signaling magnetic 
phase changes can still be defined. Next, we find that a {\em fully-connected} 
autoencoder, combined with random trees embedding produces a more or 
less temperature-resolved image of the configurations in two dimensions. 

t-SNE emerges as a clear winner among all the 
techniques, or combinations of techniques we have used, producing low-dimensional representations of data with clearly 
distinguishable patterns above and below the N\'{e}el temperature for the 3D model or the crossover temperature for the 2D model. We 
define temperature-dependent indicators and show that, in the weak-coupling regime, 
they correlate perfectly with the antiferromagntic (AFM) structure factors, and can capture a transition in the 
presence of a magnetic field. Finally, we apply t-SNE to configurations generated for the 3D 
Hubbard model away from half filling in the presence of the ``sign problem"~\cite{e_loh_90,v_iglovikov_15} in 
quantum Monte Carlo simulations and discuss the risks of 
using dimensional-reduction techniques in sign-problematic regions.

In the following section, we briefly discuss the models we have considered in this study. 
Then in Sec.~\ref{sec:methods}, we provide an overview of the various machine learning 
techniques we employ. The results are discussed in Sec.~\ref{sec:results}, followed by concluding remarks.
 
\section{Models}

\subsection{3D Ising Model}

We first consider the classical Ising model on the 3D cubic lattice
\begin{equation}
H=-J\sum_{\left<ij\right>} \sigma_i \sigma_j,
\end{equation}
where $\sigma_i=\pm1$, $\left<.. \right>$ denotes nearest neighbors, and $J$ is the strength of the corresponding exchange interaction 
(we set $J=1$ as the unit of energy whenever the Ising model is discussed). 
The system undergoes a second-order phase transition as the temperature is lowered below the critical value of $4.5J$~\cite{3DIsing}.

We perform Monte Carlo simulations based on the Metropolis algorithm~\cite{w_hastings_70} on a $N=8\times 8\times 8$ lattice 
and generate spin configurations at different temperatures. Each configuration is an array of size $N$ with $\pm 1$ as elements,
and can be thought of as a points in a $N$-dimensional space.

\subsection{The Fermi-Hubbard Model}

We are mainly interested in how quantum fluctuations affect our ability to locate phase transitions or crossovers with 
unsupervised machine learning techniques. Therefore, we also consider strongly-correlated fermions
in 3D cubic and 2D square lattices, described by the Hubbard model,
\begin{eqnarray}
H&=&-t\sum_{\left<ij\right> \sigma}c^{\dagger}_{i\sigma}
c^{\phantom{\dagger}}_{j\sigma} + U\sum_i (n_{i\uparrow}-\frac{1}{2})(n_{i\downarrow}-\frac{1}{2})\nonumber \\
&-&\mu\sum_{i\sigma} n_{i\sigma} + \frac{h}{2}\sum_i (n_{i\uparrow}-n_{i\downarrow}),
\label{eq:H}
\end{eqnarray}
where $c^{\dagger}_{i\sigma}$ ($c_{i\sigma}$) creates (annihilates) a fermion with spin $\sigma$ on site $i$, $n_{i\sigma}=c^{\dagger}_{i
\sigma} c^{\phantom{\dagger}}_{i\sigma}$ is the number operator, $t$ is the nearest neighbor hopping integral (we set $t=1$ as the unit of 
energy whenever the Hubbard model is discussed), 
$U$ is the onsite Coulomb interaction, $\mu$ is the chemical potential, and $h$ is the magnitude of the magnetic field. 
By symmetry, $\mu=0$ leads to half filling (average 
density of one fermion per site) and hole doping is achieved by decreasing $\mu$.

For any $U>0$, the 3D model displays a second-order transition from an unordered phase at high temperatures to a 
long-range N\'{e}el ordered 
phase below a $U$-dependent critical temperature $T_N$. Theoretical and numerical analysis~
\cite{r_scalettar_89,r_staudt_00,p_kent_05,t_paiva_11,e_kozik_13,d_hirschmeier_15,e_khatami_16} have shown that after an exponential 
increase from zero by turning on $U$ in the weak-coupling regime~\cite{r_scalettar_89}, $T_N$ peaks around $U=9$ and eventually goes to 
zero as $1/U$ by further increasing the interaction strength in the strong-coupling regime. The latter can be understood from the N\'{e}el 
transition of an antiferromagnetic spin$-1/2$ Heisenberg model~\cite{a_sandvik_98}, which provides the low-energy description of the 
half-filled Hubbard model in the strong-coupling regime where double occupancy is largely suppressed and fermions interact 
predominantly through the spin exchange interaction $\mathcal{J}=\frac{4t^2}{U}$. Long-range antiferromagntic correlations of the 
model have recently been observed in an experimental realization using 
ultracold fermionic atoms on optical lattices~\cite{r_hart_15}. 

We use the determinantal quantum Monte Carlo (DQMC)~\cite{r_blankenbecler_81,quest} method to simulate the model 
on a $N=4\times 4\times 4$ lattice for three 
different values of the interaction strength, $U=4$, 9, and 14, in the weak-, intermediate- and strong-coupling regimes, respectively.
We generate and save auxiliary spin configurations that can be thought of as points in a $(N\mathcal{L})$-dimensional space, 
where $\mathcal{L}=200$ is the number of imaginary 
time slices, for a range of temperatures (see Ref.~\onlinecite{k_chng_17} for details of our DQMC simulations).

We also generate auxiliary spin configurations for the model in two dimensions with $N=10\times 10$ and $U=4$
and the same number of time slices as in the 3D case. The 2D model does not have a finite-temperature 
phase transition to a long-range 
order, rather, a crossover to a region with strong antiferromagnetic correlations. The onset of this region, which is 
associated with the formation of a peak in the uniform susceptibility is estimated to be around $T=0.25$ for 
$U=4$~\cite{t_paiva_10,E_khatami_11b}.

\section{Methods}
\label{sec:methods}

\subsection{Autoencoders}

Autoencoder refers to a particular set of architectures of artificial neural networks~\cite{ANN} that can be trained to 
extract features or reduce the dimensionality of big data without the specific prior knowledge of features or distinctions 
(in an {\em unsupervised} fashion). They are made up of multiple fully-connected and/or convolutional  layers, similar to 
what is used in supervised neural network machine learning algorithms. An example of the autoencoder architecture 
is shown in the top panel of Fig.~\ref{fig:Ising}. Similarly to supervised learning, such a feed-forward network can be
trained by minimizing a cost function, which is defined based on the difference between neuron outputs at the output 
(right-most) layer and a desired output. However unlike in the supervised learning, the autoencoder is supposed to 
reproduce the input data [what is fed to the network in the input (left-most) layer] at the output layer; the desired output 
is simply the same as the input. 

The hidden layers on the left half of the autoencoder are called the {\em encoding} layers, where for example, an image is gradually 
deconstructed and represented using smaller and smaller number of pixels, whereas the hidden layers on the right half are 
called the {\em decoding} layers, where the network tries to reconstruct the image from the knowledge in the low-dimensional 
space. The middle layer, also known as the {\em coding} layer, provides the most dimensionally-reduced representation 
of the input data after the network has been trained, allowing other clustering or machine learning methods to extract 
meaningful information more easily. It is known that a single-layer autoencoder with linear activations and the PCA are 
very similar~\cite{p_baldi_89}. Here, we have used deeper fully-connected and convolutional autoencoders and have 
avoided the PCA  as it is useful only for data with linear correlations. The latter is shown to have a poor performance 
for quantum systems relative to classical ones in classifying phases~\cite{e_vanNieuwenburg_17,ScalettarBeijing}

In our convolutional autoencoder, the network tries to extract features in encoding layers by convolving a shared filter 
(or kernel), which sweeps the previous layer, with small cubic subsections on the 
data in that layer. To further reduce the 
dimensionality before the next convolutional layer, a process called {\em maxpooling} is typically performed after each 
convolutional layer in which the resolution is reduced by taking the maximum value of a subsection of data and passing 
only that value to the next layer. The decoding is done through a general process called {\em upsampling}, which refers 
to random resampling and interpolation to put together the extracted features and increase the resolution of the data. 

In fully-connected autoencoders, however, each neuron in an encoding or decoding layer is connected to all the 
neurons in the neighboring layers, as opposed to convolutional autoencoders, where 
only subsections of data from the previous layer are connected to their corresponding neurons in the following layer 
via an adjustable filter, allowing for less parameters needing to be trained. 
We use both fully-connected and convolutional autoencoders in our study using convolutional autoencoders specifically 
to allow spatially correlated features to be extracted more efficiently.

\subsection{Random Trees Embedding}

Random trees embedding transforms data in an unsupervised fashion to a high-dimensional space using tree 
graphs, resulting in sparser representation, for which the principal features can be extracted and mapped to a 
lower dimension.
Here, tree is referring to a graph with nodes repeatedly branching unidirectionally. The parent node, or the root 
of the tree, contains all the data and a node on a branch has a subset of the data. A node is associated with a 
global feature of the subset, and so, smaller branches, which contain smaller subsets farther away from the root, 
reveal more localized structures.  
Representing data in a metric space on a tree graph introduces distortions. This issue is overcome by embedding the data
using an ensemble of randomized trees instead. That is, the metric space is divided into random sections with overlapping of 
these sections permitted. A given section is then further divided up into smaller subsets on a tree. An ensemble of such trees 
can make independent observation. The maximum branching depth of a tree and the number of trees are tunable 
parameters on the algorithm

Once these random trees are grown, the pruning for features begins. The ensemble of trees votes for prominent 
features based on the density of the overlap between nodes from different trees at a given depth. Features with 
overlapping density lower than a certain threshold are discarded. This voting process reduces the bias and 
variance of those selected features. 

Here, we use random trees embedding with 100 randomized trees and a maximum depth of 10 for the 
number of branchings. We apply the algorithm to a low-dimensional representation of the auxiliary field 
data for the Hubbard model obtained via a fully-connected autoencoder as outlines in Sec.~\ref{sec:results}.

\subsection{t-SNE}

PCA has been the go-to dimensionality reduction technique in condense matter physics so far and has been 
successfully applied to classical models like the Ising model for extracting measures that closely resemble 
the order parameter or the susceptibility. However, PCA performs linear projection of the data from the 
high-dimensional space to a low-dimensional space by maximizing the variance of the projection, and so, 
local structures with non-linear correlations are not preserved. A simple measure such as the magnetization 
is evident to the linearity of the Ising model as each state can be projected onto a point in one dimension 
merely by summing its spins. The same linearity cannot be assumed for meaningful observables extracted 
from the auxiliary field configurations.

t-SNE is a powerful algorithm developed to preserve both global, and more importantly, local structures 
of  data in low-dimensional space when projecting them from a high-dimensional space.
Prior to t-SNE, SNE~\cite{g_hinton_03} was one of several attempts at achieving that. Although not very 
successful, it was the foundation for t-SNE. 
SNE employs stochastic gradient descent to minimize Kullback-Leibler divergences between pairwise 
conditional probability distributions that represent similarity of points, from the high- and low-dimensional 
spaces~\cite{tSNE_Paper}. The distributions are obtained from Gaussian
functions centered around each point. The effective number of neighbors, 
also known as the {\em perplexity}, which is provided by the user, is kept fixed by adjusting the width of 
the Gaussian distributions in different regions of the configuration space with different density of points.
t-SNE uses a slightly different cost function and Student-t distribution, as opposed to Gaussian, in the low-dimensional 
space to mitigate some issues in the original SNE and provide a better performance~\cite{tSNE_Paper}.

Values between 5 and 50 are suggested for the perplexity~\cite{tSNE_Paper,tSNE_Link}. 
We use a perplexity of about $27$, which leads to the most physically interesting 
features across various $U$'s in the low-dimensional representations of the field variables. 
The method is rather slow, scaling like $\mathcal{O}(n^{2})$, where $n$ is the number of data~\cite{z_yang_13}.

\subsection{k-means}

k-means is an unsupervised clustering algorithm that locates the centroids of k clusters in the data, where k is 
provided by the user. The algorithm starts by k centroids distributed randomly. The next step is to calculate the 
euclidean distance ($L2$-norm) of each data point to each centroid and assign them to a cluster based on which 
centroid they are closest to. The new centroids of these clusters are then calculated and these two steps are 
repeated until one converges to a stable set of centroids. This type of clustering also allows for new points to 
be classified based on which cluster they fall closest to. It is, of course, most effective in cases where data points 
form well-separated clusters. 

Here, we apply the k-means algorithm to the 2D output of our autoencoders or t-SNE to quantify the spread 
of data at various temperatures. We ask k-means to identify three centroids at each temperature. Then we find the center of the 
three centroids and define a temperature-dependent indicator, $\mathcal{D}$, which is the mean distance of 
the three centroids from their center. We note that our indicator is not unique, one may be able to 
come up with measures that more accurately capture the evolutions of features in the outputs as the temperature is varied.

\section{results}
\label{sec:results}

We start with spin configurations generated in a Monte Carlo simulation of the Ising model on a 
$N=8\times 8\times 8$ cluster. We use a fine grid for the temperature ranging from $T=3$ to $T=6$ 
in increments of 0.01 and work with a total of about 23,000 configurations across all temperatures. 
Similarly to what is done in Refs.~\cite{Wang2016,w_hu_17}, we treat each configuration as a 
point in the $N$-dimensional space and try to deduce any features corresponding to the phase 
transition by reducing the dimensionality of the configuration space and visualizing it in one or two dimensions. 

\begin{figure}[t]
\centerline {\includegraphics*[width=3.in]{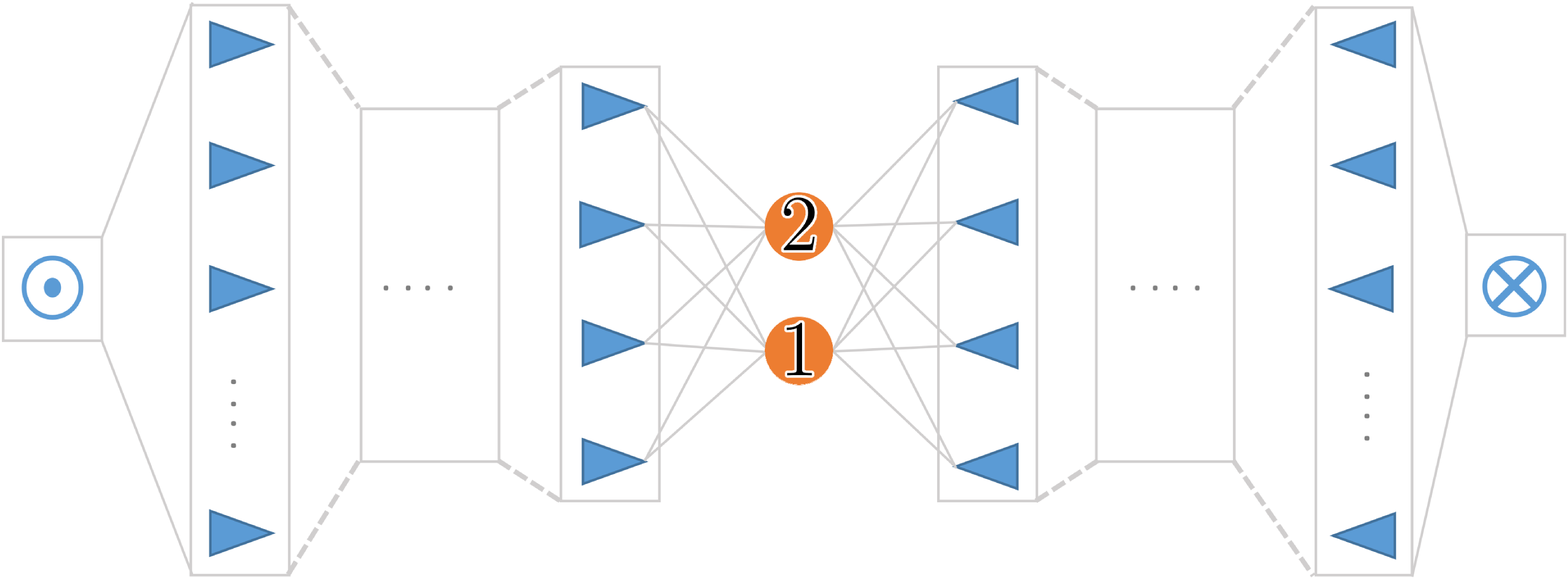}}
\centerline {\includegraphics*[width=3.in]{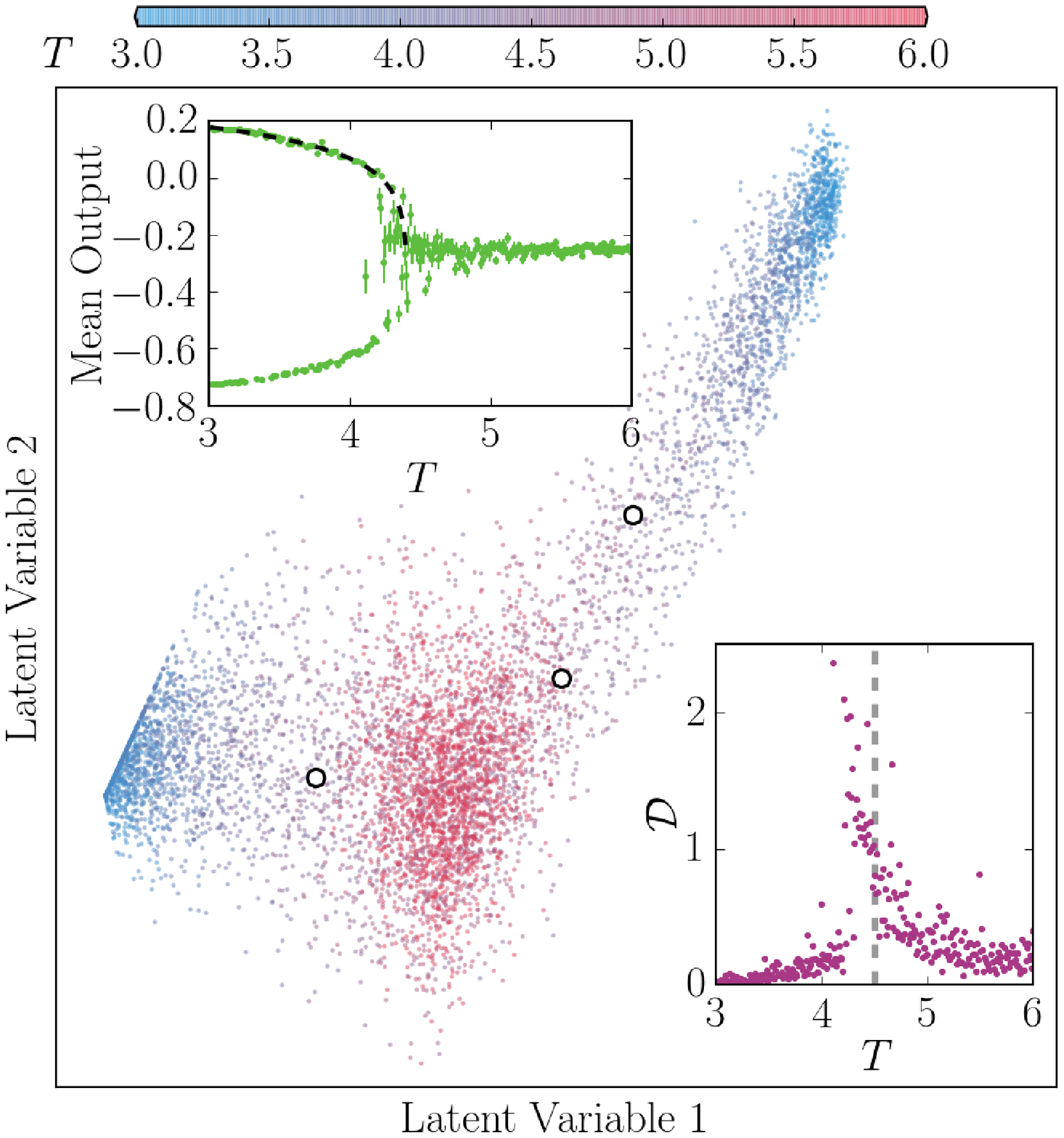}}
\caption{Values of neurons in the coding layer of a trained autoencoder that takes Monte Carlo spin configurations 
of a 3D Ising model on a $N=8\times 8\times 8$ lattice as the input. The architecture of our 3D convolutional 
autoencoder is depicted on top with neurons in the coding layer as filled (orange) circles. The encoder (decoder) 
part consists of four convolutional (upsampling) layers with 32, 8, 4, and 4 feature maps. Different symbol colors in 
the scatter plot correspond to different temperatures. {\bf Top inset:} The output of a similar autoencoder, in which 
the coding layer consists of only a single neuron, as a function of temperature. The dashed line is a fit to $A(B-T)^\beta +C$ 
with $A= 0.38$, $B=4.55$, $\beta =0.34$, and $C$ kept fixed at -0.25, which is the average output over all 
$T$. {\bf Bottom inset:} Temperature dependence of the spread of the data as measured through k-means. 
The vertical dashed lines marks the location of $T_c$.
\label{fig:Ising}}
\end{figure}

We use a 3D convolutional autoencoder with four encoding/decoding layers and one fully-connected coding 
layer with either one or two neurons. The architecture is shown in Fig.~\ref{fig:Ising}. The input/output ($\bigodot$ 
and $\bigotimes$) have the same structure as the physical system. The filter used in the convolutions is 
$2\times 2 \times 2$. A maxpooling layer is used after each convolutional layer. We shuffle the configurations 
among all temperatures, then use 70\% of them to train the neural network until the accuracy, defined as one minus the 
mean square error between the input and the output, 
saturates to a value around 75\%. We then feed the autoencoder with the remaining 30\% of the configurations 
it has not seen during the training and plot the values of neurons in the coding layer (also known as the {\em latent variables}).

The main panel in Fig.~\ref{fig:Ising} shows the output of the coding layer with two neurons. The color gradient 
of points represents the temperature gradient. There is a clear distinction between how data points cluster at 
high and low temperatures. At temperatures above the transition (red dots), we detect only one cluster. However, 
below the critical temperature, $T_c$, (blue dots) two clusters separated along both axes clearly emerge. 
This duality is a result of the spin inversion symmetry in the model that manifests itself in the ordered phase. 

At the lowest temperatures, the two clusters have the largest separation, which seems to reach a saturation 
value, analogous to the behavior of the order parameter (magnetization). However, we cannot properly quantify 
this separation as a function of temperature as the broken time reversal symmetry at a given $T<T_c$ forces 
all configurations to fall in one or the other low-temperature cluster, but not both. Having also predominantly 
one cluster of points formed at very high temperatures, suggests that in the critical temperature region, the 
points are the most spread out. We quantify the spread of data by applying the k-means clustering technique 
and requiring it to identify three clusters and their centroids (shown in Fig.~\ref{fig:Ising} at $T=4.5J$ as white 
circles). We plot $\mathcal{D}$ as a function of temperature in the bottom inset. $\mathcal{D}$ bears a striking 
resemblance to the magnetic susceptibility of the Ising model. Interestingly, thermodynamics of the system, 
encoded in the distribution of configurations in the importance sampling, are preserved during the dimension reduction.

In the top inset of Fig.~\ref{fig:Ising}, we also plot the autoencoder output in the case where we have only one 
neuron in the coding layer as a function of temperature. We observe a bifurcation of the neuron output as we 
decrease the temperature below $T_c$. Similar results were shown in Ref.~\cite{w_hu_17} for the 2D Ising model. 
The neuron output looks almost exactly like magnetization of the model as a function of temperature, except for 
a seemingly arbitrary shift. Therefore, we fit the neuron outputs in the top branch at $T<4.5J$ to a function 
proportional to $A(B-T)^\beta$ after a shift, where $A$, $B$ and $\beta$ are constants (fitted function is shown 
as a black dashed line in the inset of Fig.~\ref{fig:Ising}), and obtain  $B=4.55J$ as the critical temperature, 
which agrees well with $4.54J$, estimated for a system of the same size~\cite{3DIsing}, and $\beta=0.34$, 
close to 0.33, the critical exponent of the 3D Ising model.

Inspired by these findings, we ask if one can use a similar dimension-reduction recipe to deduce critical 
temperatures in quantum mechanical systems? In this work, we focus on magnetic phase transitions. 
We know that the perfect antiferromagnetic alignment of spins in the $z$ direction can no longer describe 
the N\'{e}el state, and so, quantum fluctuations will likely blur the clear image we observe in reduced 
dimensions for the classical Ising model. However, to what extent can the information be still useful? 

We set to answer this question by using the auxiliary fields and  
modifying our convolutional autoencoder such that the $\mathcal{L}=200$ imaginary time slices are treated as different 
``color" channels in the input/output layers ($\mathcal{L}$ $\bigodot$ and $\bigotimes$), each with $L\times L\times L$ neurons, 
the same as the single channel we used for the Ising model. The architecture of the remaining network is also modified to 
suite the smaller $L$ ($=4$) we use for the Hubbard model. We  use one less hidden layer and different number of feature 
maps in hidden layers than in the Ising autoencoder. The output in this case is sensitive to the number of feature maps and 
we have chosen a set that results in the largest accuracy.

\begin{figure}[t]
\centerline {\includegraphics*[width=3.3in]{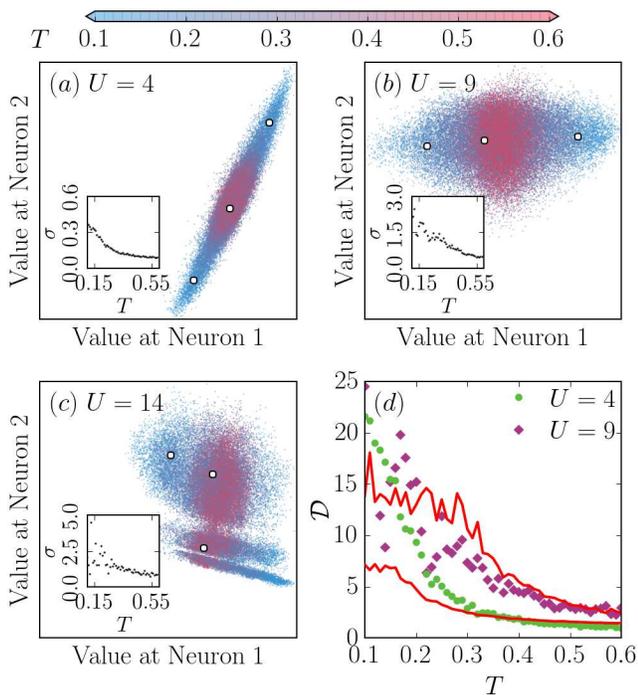}}
\caption{The output of the 3D convolutional autoencoder for the 3D Hubbard model with (a) $U=4$, (2) $9$, 
and (3) $U=14$ at half filling. Each of the $\mathcal{L}=200$ imaginary time slices are treated as a different 
``color" channel. The input is a four-dimensional array of size $\mathcal{L}\times N$ with $N=4^{3}$. Similarly to 
the output for the Ising model, the character of the output in the space of the neuron outputs of the coding layer 
changes as a function of temperature. However, we do not find a formation of distinct clusters below the expected 
N\'{e}el temperatures. In (d), we show $\mathcal{D}$ for the data in (a) and (b), normalized to fit the antiferromagnetic 
structure factors (shown as red lines) in each case, as a function of temperature. For $U=14$, this indicator is 
dominated by noise, and therefore, not shown.
\label{fig:Hubb2D}}
\end{figure}

The results are shown in Fig.~\ref{fig:Hubb2D} for $U=4$, 9 , and 14. The the first three main panels show the 
outcome in the case where the coding layer has two neurons. Unlike the classical case, lowering the temperature 
does not lead to the formation and contraction of two distinguishable clusters. Instead, the data points seem to 
spread out mostly along one of the two dimensions, at least for $U=4$ and $U=9$. Quantum fluctuations and the 
fact that our auxiliary field is represented along a particular direction in the spin vector space seem to play a 
significant role in blurring the picture. As a result, in this case, our measure of the spread of data obtained from 
k-means, $\mathcal{D}$, behaves more similarly to an order parameter than to the susceptibility. In Fig.~\ref{fig:Hubb2D}(d) 
we show this quantity, along with the AFM structure factor~\cite{AFMSF}, calculated in the DQMC, as a function 
of temperature for $U=4$ and 9.  $\mathcal{D}$ has been multiplied by a constant that minimizes the mean square 
distance of its values from the structure factor data (weighted by the error bars in the latter) over all temperatures. 
We do not find a good agreement between the two at low temperatures, however, $\mathcal{D}$ is a relatively 
smooth function for $U=4$ and displays the fastest rise around $T=0.2$, where we expect the critical temperature 
to be for this cluster. For $U=9$, we observe large fluctuations in $\mathcal{D}$, which appear to grow significantly 
larger below $T= 0.35$, the expected Ne\'{e}l temperature. For $U=14$, the indicator is dominated by noise and is 
not shown. Our approach seems to be most efficient in locating the critical behavior in the weak coupling regime 
of the Hubbard model.

The same conclusion can be drawn considering the output of an autoencoder that has a single neuron in the coding layer. 
Unlike for the Ising model, we do not find a bifurcation as a function of temperature when projecting the data to one dimension. 
However, we find that the fluctuations in the data increase rapidly as the temperature decreases. So, in the insets of  
Figs.~\ref{fig:Hubb2D}(a),  \ref{fig:Hubb2D}(b) and \ref{fig:Hubb2D}(c), we show the standard deviation of the latent variable as a function of 
temperature. For $U=4$ and $U=9$, it behaves similarly to $\mathcal{D}$, e.g., rapidly rises around $T_N$ for $U=4$. 
For $U=14$, the fluctuation dominate at $T<0.3$.

\begin{figure}[t]
\centerline {\includegraphics*[width=2.7in]{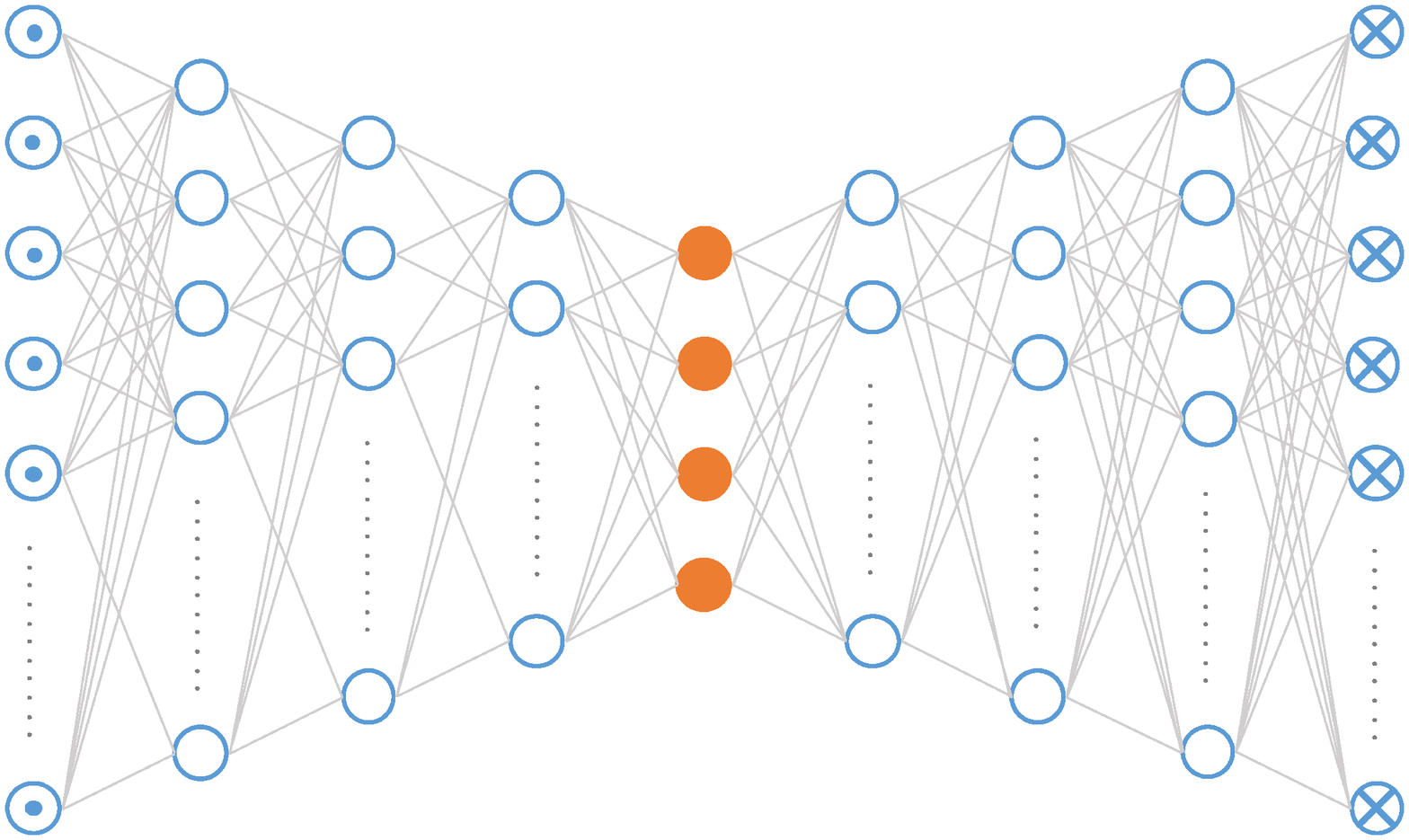}}
\centerline {\includegraphics*[width=3.3in]{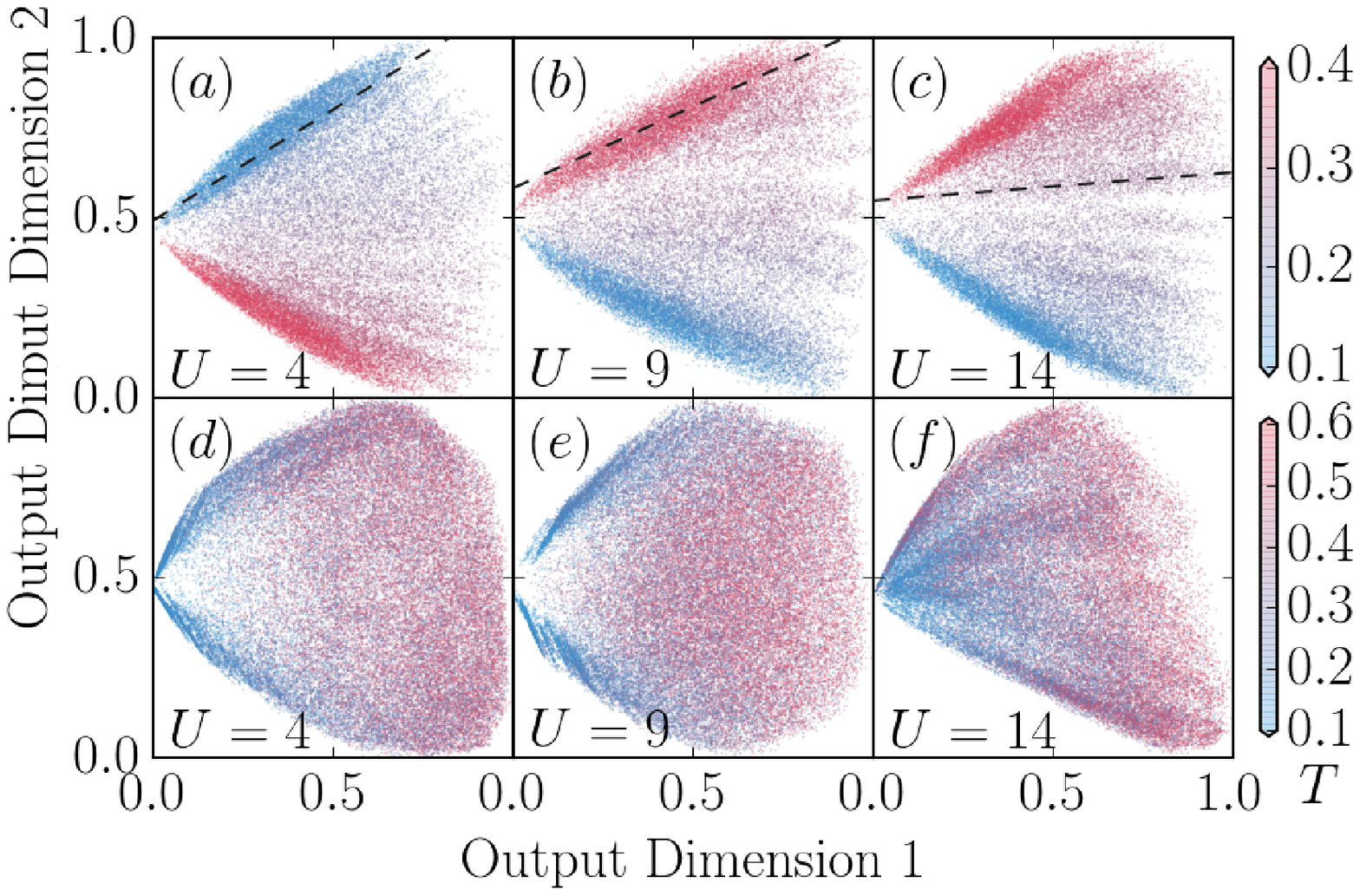}}
\caption{The output of random trees embedding algorithm trained on the four latent variables of a fully-connected autoencoder for the 3D 
Hubbard model with (a) $U=4$, (b) $U=9$, and (c) $U=14$ at half filling. The input layer of the autoencoder (shown on top)
is a 1D array of size $\mathcal{L}\times N$. The hidden layers in the encoder/decoder parts have 80, 30, and 10 neurons. The 
random trees embedding algorithm further reduces the dimension of the data from four to two. In this case, the outputs clearly separate data points from different temperatures. The dashed lines are line fits to data at the estimated N\'{e}el temperature for each $U$. (d)-(f) Same as in (a)-(c), except that the input are the latent variables of the convolutional autoencoder used for Fig.~\ref{fig:Hubb2D}, modified to have four neurons in the coding layer. In this case, the temperature gradient is larger along output 1 and clustering of points at low temperatures is visible for $U=4$ and $U=9$.
\label{fig:RandomTrees}}
\end{figure}

Next, we explore the possibility of having more than two neurons in the coding layer (more than two latent variables) 
and further reducing the dimensionality of data using another technique such as random 
trees embedding. We use the same convolutional autoencoder architecture as we used for Fig.~\ref{fig:Hubb2D}, except that 
we choose to have four neurons in the coding layer. Then, we feed the output of the autoencoder to the random trees 
embedding algorithm and obtain a representation in two dimensions. The results are summarized in Fig.~\ref{fig:RandomTrees}(d)-(f). 
The emerging fan shape not only creates an approximate temperature resolution, but for $U=4$ and 9 also 
exhibits two low-temperature clusters near 
small values along the first dimension, separated in along second dimension, reminiscent of the Ising picture. For $U=14$, the latter
feature is mostly washed away. 

The convolutions in our autoencoder appear to be crucial for the low-temperature clustering in the output 
of the random trees embedding, but not 
for the temperature resolution. We try the same combination of autoencoders and random trees embedding, 
except that this time, we use a fully-connected autoencoder, shown in the top part of Fig.~\ref{fig:RandomTrees}. 
The results for the latter are shown in Fig.~\ref{fig:RandomTrees}(a)-(c). We find that despite the better temperature
resolution than with the convolutional autoencoder, which extends even to $U=14$, 
there is no peculiarity that can point to different phases at high and low temperatures. 
Interestingly, data points corresponding to a given temperature spread along straight lines in this reduced space with the 
slope of the line correlating with temperature.
We have also explored applying random trees embedding to data extracted not from the coding layer, but from various hidden 
layers of our fully-connected autoencoder. We find that the coding layer yields the best picture in terms of temperature resolution.

\begin{figure}[t]
\centerline {\includegraphics*[width=3.3in]{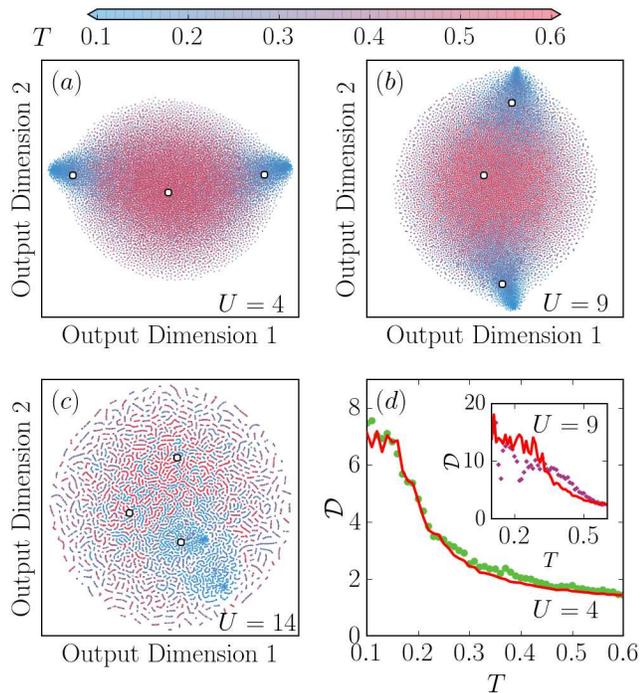}}
\caption{Output of the t-SNE algorithm in two dimensions for the 3D Hubbard model at half filling for three 
different values of the interaction strength, (a) $U=4$, (b) $U=9$, and (c) $U=14$. We use 
the raw auxiliary field configurations from DQMC simulations of the model at 51 temperatures on a 
uniform grid that extends from $T=0.10$ to $T=0.60$ as input. Different symbol colors in (a)-(c) 
correspond to different temperatures. (d) Same indicator as in Fig.~\ref{fig:Hubb2D}(d) as a 
function of temperature calculated for the t-SNE outputs. For $U=4$, the indicator follows the AFM structure factor very closely.
\label{fig:tsne}}
\end{figure}

Techniques like random trees embedding do not scale well with the dimension of the original configuration 
space and so cannot easily be directly applied to the raw auxiliary field configurations for dimension reduction. 
Other clustering techniques, such as the PCA or the t-SNE, can better handle larger dimensions. As we discuss 
below, we find that the direct application of t-SNE to the raw data also yields a superior distinction between 
clustering patterns at different temperatures.

We apply the t-SNE algorithm~\cite{tSNE_Code} with perplexity of 27.07 to the half-filled 3D Hubbard configurations at our three 
interaction strengths. We reduce the dimensionality to 37 after preprocessing using PCA by a batch size 
of 500 data at a time within the t-SNE algorithm for greater speed without inducing severe distortions and 
to filter out  some noise~\cite{tSNE_Paper}. These are the same configuration as the ones we used for the autoencoder. 
The two-dimensional visualizations are shown in Figs.~\ref{fig:tsne}. 
For $U=4$ in Fig.~\ref{fig:tsne}(a), not only the data points spread out by decreasing the temperature, 
similarly to the autoencoder outcome in Fig.~\ref{fig:Hubb2D}, but also two distinct clusters emerge, 
analogous to what we observed in Fig.~\ref{fig:RandomTrees} from random trees embedding, or for 
the 3D Ising model. We point out, however, that unlike in the Ising case, there is a significant number 
of points that are scattered between the two centers at the lowest temperatures.

A similar picture emerges for $U=9$ in Fig.~\ref{fig:tsne}(b), however for $U=14$, we find that the 
data points stick together and form worm-like figures. 
As can be seen in Fig.~\ref{fig:tsne}(c), they mostly gather around two centers at low temperatures. 
Interestingly, we find that they are formed by data points that belong to the same temperature to a great extent.
We attribute their formation to Mott physics. In the Mott region at large $U$, the freezing of charge 
degrees of freedom manifests itself in a significant increase in the autocorrelation time in the single spin flip scheme of our 
DQMC algorithm, and the simulations become less ergodic than in the weak- or intermediate-coupling regimes.
We observe this behavior despite the fact that we have attempted to mitigate the problem by performing 10 
different simulations of the model for $U=14$ using different random number seeds and shuffling the configurations
from those simulations before applying t-SNE.

We find a remarkable correlation of the indicator $\mathcal{D}$, calculated for the 2D t-SNE outputs, and 
the AFM structure factor of the model in the weak-coupling regime. The two are plotted in Fig.~\ref{fig:tsne}(d)
for $U=4$ after normalizing $\mathcal{D}$ to best fit the structure factor (red solid curve). They show a very good 
agreement across the entire range of temperatures shown. The development of long-range correlations and 
the growing dissimilarities in the configurations due to the breaking of the time-reversal symmetry as we lower 
the temperature can be directly mapped to the increase in the structure factor. 

We have also applied t-SNE to the auxiliary spin conÞgurations for visualizations in 1D and 3D (not shown). 
We do not find any meaningful features in the 1D visualization. Projection of the configuration to 3D produces 
outputs that resemble volumetric versions of the 2D scatter plots. 

\begin{figure}[t]
\centerline {\includegraphics*[width=3.3in]{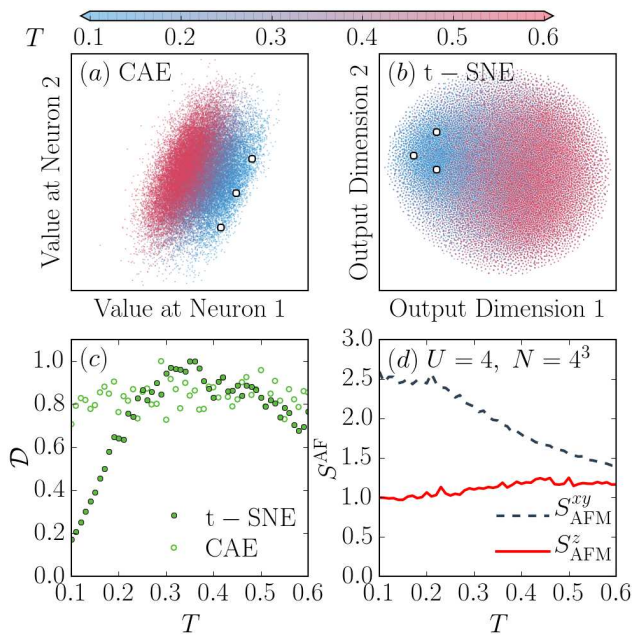}}
\caption{(a) The convolutional autoencoder, and (b) t-SNE outputs for the 3D Hubbard model with $U=4$ at half filling in the 
presence of a magnetic field $h=1.0$. (c) The corresponding indicators as a function of temperature. (d) 
The AFM structure factors of the model for the $z$ and $xy$ components of the spin as a function of temperature.
\label{fig:Mu1.0}}
\end{figure}

Breaking the SU(2) symmetry of the Hubbard model changes the low-temperature physics and the picture obtained 
from unsupervised machine learning algorithms. We explore that by including a strong uniform magnetic field, 
$h=1.0$, which mostly aligns the spins with the $z$ axis and pushes the remnant AFM correlations to the $xy$ plane. 
The canted AFM physics for the 3D Hubbard model is evidenced by the $z$ and $xy$ components of the AFM structure factor
plotted in Fig.~\ref{fig:Mu1.0}(d). The ferromagnetic correlations along $z$ (not shown) are more than a factor of two larger 
than the $xy$ component of the AFM correlations. 

We study the effect of the magnetic field on the results from the convolutional 
autoencoder and the t-SNE. The output of the autoencoder with two latent variables, shown in Fig.~\ref{fig:Mu1.0}(a), 
displays a temperature resolution, however, unlike for the original model with SU(2) symmetry, all of the low-temperature 
points appear to belong to a single cluster. As shown in Fig.~\ref{fig:Mu1.0}(c), $\mathcal{D}$ for this case is a flat function of
temperature, which misses the important phase changes in the $xy$ plane. One can in principle define other indicators. For
example, the distance between the high-temperature 
cluster and clusters at lower temperatures can presumably serve as a measure for the ferromagnetic correlations along $z$.
The t-SNE, on the other hand,  
in addition to displaying the temperature resolution, assignes low-temperature points to a region that shrinks 
rapidly by lowering the temperature below $T\sim 0.3$, signaling a phase change similar to the case of zero magnetic field. 
This is more clearly captured by the corresponding indicator in Fig.~\ref{fig:Mu1.0}(c), which now measures 
simply the concentration of points.

\begin{figure}[t]
\centerline {\includegraphics*[width=3.3in]{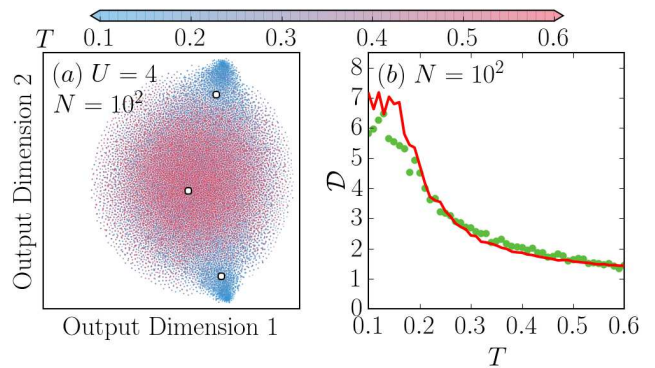}}
\caption{Output of the t-SNE algorithm for the 2D Hubbard model with $N=10^{2}$ and $U=4$ at half filing. 
We use 800 raw auxiliary field configurations for each temperature in the uniform grid of $T$ between 0.1 and 0.60. 
We use the same t-SNE parameters as in the training for the 3D Hubbard model in Fig.~\ref{fig:tsne}. 
(b) The indicator $\mathcal{D}$ (green symbols) and the AFM structure factor (red line) as a function of temperature.}
\label{fig:tsne2DHubb}
\end{figure}

Motivated by the ability of t-SNE to distinguish high-temperature configurations from the low-temperature ones 
below the critical temperatures of the 3D Hubbard model, at least for small $U$, we examine the 2D Hubbard 
model in the weak-coupling regime, also at half filling, using t-SNE with the same parameters as used for the 3D model. 
As shown in Fig.~\ref{fig:tsne2DHubb}, the outcome of t-SNE in this case exhibits a similar extension of points in the 
space by lowering the temperature as for the model in 3D. We do not find any specific features that point to a 
crossover, as opposed to a phase transition expected in 3D. 
In fact, the indicator $\mathcal{D}$, which closely follows the AFM structure factor in 
this case too, exhibits an even sharper rise just below $T=0.2$ than for the model in 3D.

\begin{figure}[t]
\centerline {\includegraphics*[width=3.3in]{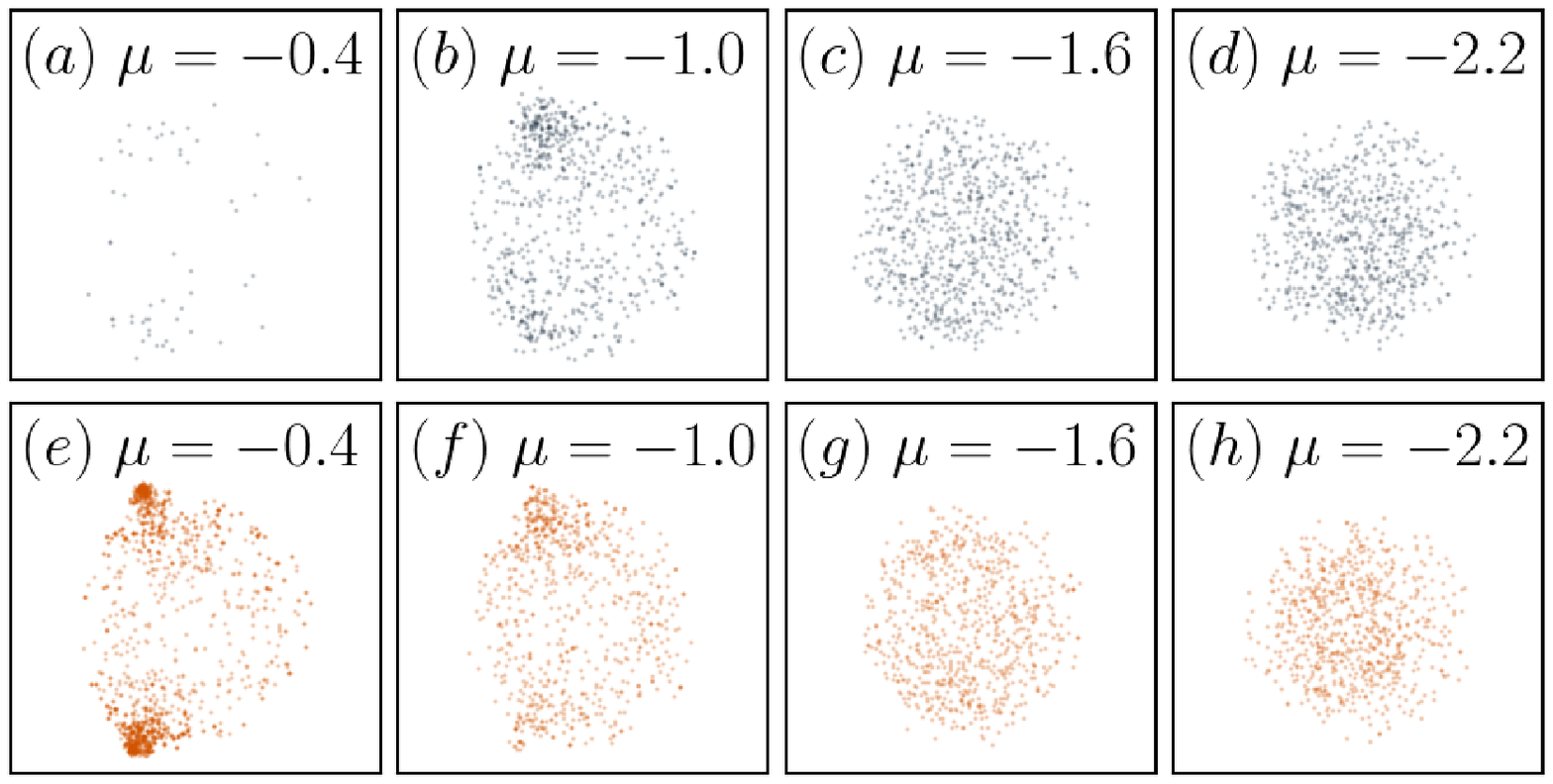}}
\centerline {\includegraphics*[width=3.3in]{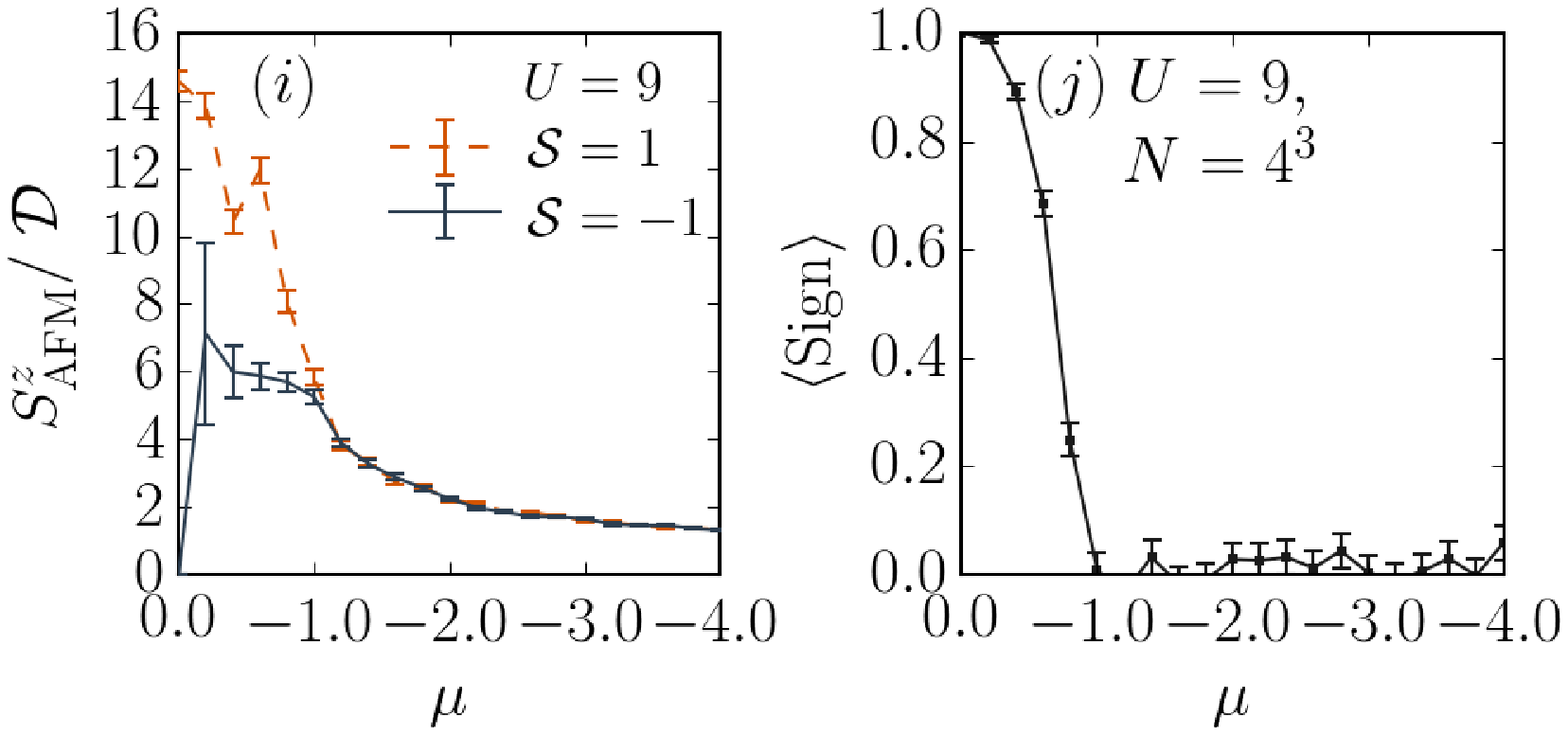}}
\caption{(a)-(h) The t-SNE output for the 3D Hubbard model with $U=9$ away from half-filing. We have used 1,500 auxiliary field 
configurations per chemical potential $\mu$, generated deep in the AFM phase ($T=0.16$) for $\mu$ ranging from 
$\mu$=-0.2 to $\mu$=-4.0 in steps of 0.2. We show snapshots of the output at select $\mu$, separated into configurations with 
negative sign ($\mathcal{S}=-1$) in the top row and those with a positive sign ($\mathcal{S}=+1$) in the bottom row. We use the 
same t-SNE parameters as in the training for the 3D Hubbard model at half filling. (i) The expectation value of the AFM structure 
factor obtained using configurations with positive or negative signs separately as a function of the chemical potential. 
(j) The average sign as a function of $\mu$.}
\label{fig:away}
\end{figure}

One may be tempted to apply these techniques to configurations generated for the Fermi-Hubbard 
model away from half filling to, for example, study the fate of the antiferromagnetic phase of the model in 3D, 
or the pseudogap and superconducting properties in 2D, as the system is doped. However, any phase 
transition or crossover deduced from the dimensionally-reduced data in that case, for example, through 
an indicator similar to $\mathcal{D}$, would be a transition or crossover not for the Fermi-Hubbard model, 
but for an alternative model whose statistics is described by the absolute
value of the probability amplitudes in the DQMC simulation of the Fermi-Hubbard model. Therefore, unless 
it so happens that the phase boundaries of the two models are the same for the transition/crossover of 
interest, the results will not be of much use for the Fermi-Hubbard model.

Here, we demonstrate this concept using DQMC simulations of the 3D Hubbard model for $U=9$ at 
nonzero hole doping ($\mu<0$). We generate 2000 auxiliary spin configurations per $\mu$ that ranges 
from $-0.2$ to $-4.0$ in increments of 0.2 at $T=0.16$, which is deep in the AFM phase at half filling. We first ignore the 
signs of the configurations and run t-SNE on 3/4 of the entire set. Then, in our visualization of the 2D output, 
we separate configurations with a negative sign ($\mathcal{S}=-1$) from those with a
positive sign ($\mathcal{S}=1$) at different values of $\mu$. The results are shown in 
Fig.~\ref{fig:away}(a)-(d) and Fig.~\ref{fig:away}(e)-(h), respectively. We have seen an indicator can be 
defined based on the t-SNE output that behaves very similarly to a physical observable (e.g., the structure factor), 
although this was not exactly the case for our $\mathcal{D}$ at $U=9$.
If we assume this is valid away from half filling and separately for the positive and negative 
configurations, the indicators are expected to correlate with the corresponding structure factors 
calculated using positive only or negative only configurations [see Fig.~\ref{fig:away}(i)].
However, the physical structure factor in the presence of the sign problem in the DQMC is obtained from the 
difference between the above two structure factors divided by the average sign [the latter is shown in Fig.~\ref{fig:away}(j)
as a function of $\mu$]. Therefore, one cannot expect an indicator, when calculated using a mixture of negative 
and positive configurations at every $\mu$ (effectively ignoring the sign) to yield any physically meaningful quantity.
This is specially evident in our case at $\mu<-1.0$, where the sign problem is severe and the scatter plots for 
$\mathcal{S}=-1$ and $\mathcal{S}=+1$ look essentially alike.

\begin{figure}[t]
\centerline {\includegraphics*[width=3.3in]{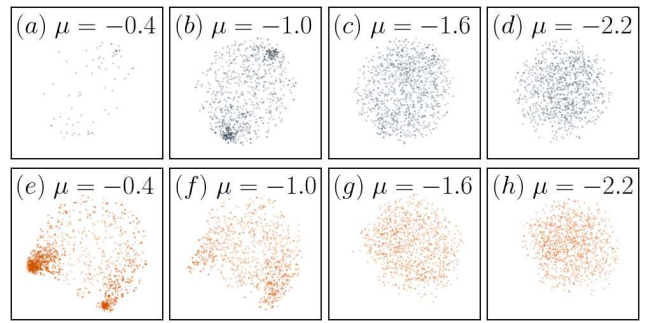}}
\caption{Same as Fig.~\ref{fig:away}, except that here, t-SNE calculations have been performed separately 
on configurations with the same sign. We also use 2,000 auxiliary field configurations per $\mu$.}
\label{fig:away2}
\end{figure}

Finally, we examine the application of t-SNE separately on configurations with positive or negative signs to make
sure that the results in Fig.~\ref{fig:away} are not biased due to the dominance of the $\mathcal{S}=1$ configurations
near half filling. Figure~\ref{fig:away2} shows the same trend in the similarity of the scatter plots between the configurations
with different signs when the average sign drops to zero.

In summary, we have applied various unsupervised machine learning techniques, such as autoencoders, random trees 
embedding, k-means, and t-SNE, to obtain low-dimensional representations for the auxiliary spin configurations of the 
Fermi-Hubbard models in different interaction regimes. We show that one can extract features from the data in reduced 
dimensions that resemble physical observables related to the magnetic correlations in the physical system. 
The configurations are sampled during DQMC simulations at finite temperatures at and away from the half-filing. 

As a benchmark, we first train a convolutional autoencoder using spin configurations of a 3D classical 
Ising model and obtain indicators that closely resemble magnetization and susceptibility. The low-dimensional 
representations of autoencoders, or a combination of them with random trees embedding techniques, 
however, are largely affected by quantum fluctuations in the Hubbard model,
preventing us from mapping the results to physical observables despite the existence of distinct 
features that can point to a phase transition at low temperatures at least in the weak-coupling regime. 

On the other hand, we find that the t-SNE algorithm combined with k-means provides results that 
perfectly correlate with the AFM structure factor of the model in two or three spatial dimensions 
as a function of temperature. 
We also explore the effect of a symmetry breaking magnetic field on the outcome of the unsupervised 
techniques and show that t-SNE is also capable of capturing the transition to a different type of phase at 
low temperatures. We employ t-SNE to demonstrate that the measures we extract from 
low-dimensional representations of the auxiliary fields in order to describe the physics can no longer serve 
that purpose away from half filling in the presence of the sign problem in the DQMC simulations.

We thank Demetrius Almada for providing the Monte Carlo data for the 3D Ising model. We acknowledge 
support from the National Science Foundation (NSF) under the Grant No. DMR-1609560. The computations 
were performed in part on the Teal computer cluster of the Department of Physics and Astronomy of San 
Jose State University and in part on the Spartan high-performance computing facility at San Jose 
State University supported by the NSF under the Grant No. OAC-1626645.


\begin{thebibliography}{52}%
\makeatletter
\providecommand \@ifxundefined [1]{%
 \@ifx{#1\undefined}
}%
\providecommand \@ifnum [1]{%
 \ifnum #1\expandafter \@firstoftwo
 \else \expandafter \@secondoftwo
 \fi
}%
\providecommand \@ifx [1]{%
 \ifx #1\expandafter \@firstoftwo
 \else \expandafter \@secondoftwo
 \fi
}%
\providecommand \natexlab [1]{#1}%
\providecommand \enquote  [1]{``#1''}%
\providecommand \bibnamefont  [1]{#1}%
\providecommand \bibfnamefont [1]{#1}%
\providecommand \citenamefont [1]{#1}%
\providecommand \href@noop [0]{\@secondoftwo}%
\providecommand \href [0]{\begingroup \@sanitize@url \@href}%
\providecommand \@href[1]{\@@startlink{#1}\@@href}%
\providecommand \@@href[1]{\endgroup#1\@@endlink}%
\providecommand \@sanitize@url [0]{\catcode `\\12\catcode `\$12\catcode
  `\&12\catcode `\#12\catcode `\^12\catcode `\_12\catcode `\%12\relax}%
\providecommand \@@startlink[1]{}%
\providecommand \@@endlink[0]{}%
\providecommand \url  [0]{\begingroup\@sanitize@url \@url }%
\providecommand \@url [1]{\endgroup\@href {#1}{\urlprefix }}%
\providecommand \urlprefix  [0]{URL }%
\providecommand \Eprint [0]{\href }%
\providecommand \doibase [0]{http://dx.doi.org/}%
\providecommand \selectlanguage [0]{\@gobble}%
\providecommand \bibinfo  [0]{\@secondoftwo}%
\providecommand \bibfield  [0]{\@secondoftwo}%
\providecommand \translation [1]{[#1]}%
\providecommand \BibitemOpen [0]{}%
\providecommand \bibitemStop [0]{}%
\providecommand \bibitemNoStop [0]{.\EOS\space}%
\providecommand \EOS [0]{\spacefactor3000\relax}%
\providecommand \BibitemShut  [1]{\csname bibitem#1\endcsname}%
\let\auto@bib@innerbib\@empty
\bibitem [{\citenamefont {Arsenault}\ \emph {et~al.}(2014)\citenamefont
  {Arsenault}, \citenamefont {Lopez-Bezanilla}, \citenamefont {von
  Lilienfeld},\ and\ \citenamefont {Millis}}]{l_arsenault_14}%
  \BibitemOpen
  \bibfield  {author} {\bibinfo {author} {\bibfnamefont {Louis-Fran\c{c}ois}\
  \bibnamefont {Arsenault}}, \bibinfo {author} {\bibfnamefont {Alejandro}\
  \bibnamefont {Lopez-Bezanilla}}, \bibinfo {author} {\bibfnamefont
  {O.~Anatole}\ \bibnamefont {von Lilienfeld}}, \ and\ \bibinfo {author}
  {\bibfnamefont {Andrew~J.}\ \bibnamefont {Millis}},\ }\bibfield  {title}
  {\enquote {\bibinfo {title} {Machine learning for many-body physics: The case
  of the anderson impurity model},}\ }\href {\doibase
  10.1103/PhysRevB.90.155136} {\bibfield  {journal} {\bibinfo  {journal} {Phys.
  Rev. B}\ }\textbf {\bibinfo {volume} {90}},\ \bibinfo {pages} {155136}
  (\bibinfo {year} {2014})}\BibitemShut {NoStop}%
\bibitem [{\citenamefont {Carrasquilla}\ and\ \citenamefont
  {Melko}(2017)}]{j_carrasquilla_16}%
  \BibitemOpen
  \bibfield  {author} {\bibinfo {author} {\bibfnamefont {Juan}\ \bibnamefont
  {Carrasquilla}}\ and\ \bibinfo {author} {\bibfnamefont {Roger~G.}\
  \bibnamefont {Melko}},\ }\bibfield  {title} {\enquote {\bibinfo {title}
  {Machine learning phases of matter},}\ }\href
  {http://dx.doi.org/10.1038/nphys4035} {\bibfield  {journal} {\bibinfo
  {journal} {Nat Phys}\ }\textbf {\bibinfo {volume} {13}},\ \bibinfo {pages}
  {431--434} (\bibinfo {year} {2017})},\ \bibinfo {note} {letter}\BibitemShut
  {NoStop}%
\bibitem [{\citenamefont {van Nieuwenburg}\ \emph {et~al.}(2017)\citenamefont
  {van Nieuwenburg}, \citenamefont {Liu},\ and\ \citenamefont
  {Huber}}]{e_vanNieuwenburg_17}%
  \BibitemOpen
  \bibfield  {author} {\bibinfo {author} {\bibfnamefont {Evert P.~L.}\
  \bibnamefont {van Nieuwenburg}}, \bibinfo {author} {\bibfnamefont {Ye-Hua}\
  \bibnamefont {Liu}}, \ and\ \bibinfo {author} {\bibfnamefont {Sebastian~D.}\
  \bibnamefont {Huber}},\ }\bibfield  {title} {\enquote {\bibinfo {title}
  {Learning phase transitions by confusion},}\ }\href
  {http://dx.doi.org/10.1038/nphys4037} {\bibfield  {journal} {\bibinfo
  {journal} {Nat Phys}\ }\textbf {\bibinfo {volume} {13}},\ \bibinfo {pages}
  {435--439} (\bibinfo {year} {2017})},\ \bibinfo {note} {letter}\BibitemShut
  {NoStop}%
\bibitem [{\citenamefont {Broecker}\ \emph {et~al.}(2016)\citenamefont
  {Broecker}, \citenamefont {Carrasquilla}, \citenamefont {Melko},\ and\
  \citenamefont {Trebst}}]{p_broecker_16}%
  \BibitemOpen
  \bibfield  {author} {\bibinfo {author} {\bibfnamefont {Peter}\ \bibnamefont
  {Broecker}}, \bibinfo {author} {\bibfnamefont {Juan}\ \bibnamefont
  {Carrasquilla}}, \bibinfo {author} {\bibfnamefont {Roger~G.}\ \bibnamefont
  {Melko}}, \ and\ \bibinfo {author} {\bibfnamefont {Simon}\ \bibnamefont
  {Trebst}},\ }\bibfield  {title} {\enquote {\bibinfo {title} {Machine learning
  quantum phases of matter beyond the fermion sign problem},}\ }\href
  {https://arxiv.org/pdf/1608.07848.pdf} {\  (\bibinfo {year} {2016})},\
  \Eprint {http://arxiv.org/abs/arXiv:1608.07848} {arXiv:1608.07848}
  \BibitemShut {NoStop}%
\bibitem [{\citenamefont {Ch'ng}\ \emph {et~al.}(2017)\citenamefont {Ch'ng},
  \citenamefont {Carrasquilla}, \citenamefont {Melko},\ and\ \citenamefont
  {Khatami}}]{k_chng_17}%
  \BibitemOpen
  \bibfield  {author} {\bibinfo {author} {\bibfnamefont {Kelvin}\ \bibnamefont
  {Ch'ng}}, \bibinfo {author} {\bibfnamefont {Juan}\ \bibnamefont
  {Carrasquilla}}, \bibinfo {author} {\bibfnamefont {Roger~G.}\ \bibnamefont
  {Melko}}, \ and\ \bibinfo {author} {\bibfnamefont {Ehsan}\ \bibnamefont
  {Khatami}},\ }\bibfield  {title} {\enquote {\bibinfo {title} {Machine
  learning quantum phases of strongly-correlated fermions},}\ }\href
  {https://arxiv.org/abs/1609.02552} {\bibfield  {journal} {\bibinfo  {journal}
  {To appear in Phys. Rev. X}\ } (\bibinfo {year} {2017})},\ \Eprint
  {http://arxiv.org/abs/arXiv:1609.02552} {arXiv:1609.02552} \BibitemShut
  {NoStop}%
\bibitem [{\citenamefont {Zhang}\ and\ \citenamefont
  {Kim}(2017)}]{y_zhang_17a}%
  \BibitemOpen
  \bibfield  {author} {\bibinfo {author} {\bibfnamefont {Yi}~\bibnamefont
  {Zhang}}\ and\ \bibinfo {author} {\bibfnamefont {Eun-Ah}\ \bibnamefont
  {Kim}},\ }\bibfield  {title} {\enquote {\bibinfo {title} {Quantum loop
  topography for machine learning},}\ }\href {\doibase
  10.1103/PhysRevLett.118.216401} {\bibfield  {journal} {\bibinfo  {journal}
  {Phys. Rev. Lett.}\ }\textbf {\bibinfo {volume} {118}},\ \bibinfo {pages}
  {216401} (\bibinfo {year} {2017})}\BibitemShut {NoStop}%
\bibitem [{\citenamefont {Zhang}\ \emph {et~al.}(2017)\citenamefont {Zhang},
  \citenamefont {Melko},\ and\ \citenamefont {Kim}}]{y_zhang_17b}%
  \BibitemOpen
  \bibfield  {author} {\bibinfo {author} {\bibfnamefont {Yi}~\bibnamefont
  {Zhang}}, \bibinfo {author} {\bibfnamefont {Roger~G.}\ \bibnamefont {Melko}},
  \ and\ \bibinfo {author} {\bibfnamefont {Eun-Ah}\ \bibnamefont {Kim}},\
  }\bibfield  {title} {\enquote {\bibinfo {title} {Machine learning z$_2$
  quantum spin liquids with quasi-particle statistics},}\ }\href
  {https://arxiv.org/pdf/1705.01947.pdf} {\  (\bibinfo {year} {2017})},\
  \Eprint {http://arxiv.org/abs/arXiv:1705.01947} {arXiv:1705.01947}
  \BibitemShut {NoStop}%
\bibitem [{\citenamefont {Torlai}\ and\ \citenamefont
  {Melko}(2016)}]{Torlai2016}%
  \BibitemOpen
  \bibfield  {author} {\bibinfo {author} {\bibfnamefont {Giacomo}\ \bibnamefont
  {Torlai}}\ and\ \bibinfo {author} {\bibfnamefont {Roger~G.}\ \bibnamefont
  {Melko}},\ }\bibfield  {title} {\enquote {\bibinfo {title} {Learning
  thermodynamics with boltzmann machines},}\ }\href {\doibase
  10.1103/PhysRevB.94.165134} {\bibfield  {journal} {\bibinfo  {journal} {Phys.
  Rev. B}\ }\textbf {\bibinfo {volume} {94}},\ \bibinfo {pages} {165134}
  (\bibinfo {year} {2016})}\BibitemShut {NoStop}%
\bibitem [{\citenamefont {Carleo}\ and\ \citenamefont
  {Troyer}(2017)}]{Carleo2016}%
  \BibitemOpen
  \bibfield  {author} {\bibinfo {author} {\bibfnamefont {Giuseppe}\
  \bibnamefont {Carleo}}\ and\ \bibinfo {author} {\bibfnamefont {Matthias}\
  \bibnamefont {Troyer}},\ }\bibfield  {title} {\enquote {\bibinfo {title}
  {Solving the quantum many-body problem with artificial neural networks},}\
  }\href {\doibase 10.1126/science.aag2302} {\bibfield  {journal} {\bibinfo
  {journal} {Science}\ }\textbf {\bibinfo {volume} {355}},\ \bibinfo {pages}
  {602--606} (\bibinfo {year} {2017})}\BibitemShut {NoStop}%
\bibitem [{\citenamefont {Torlai}\ \emph {et~al.}(2017)\citenamefont {Torlai},
  \citenamefont {Mazzola}, \citenamefont {Carrasquilla}, \citenamefont
  {Troyer}, \citenamefont {Melko},\ and\ \citenamefont {Carleo}}]{g_torlai_17}%
  \BibitemOpen
  \bibfield  {author} {\bibinfo {author} {\bibfnamefont {Giacomo}\ \bibnamefont
  {Torlai}}, \bibinfo {author} {\bibfnamefont {Guglielmo}\ \bibnamefont
  {Mazzola}}, \bibinfo {author} {\bibfnamefont {Juan}\ \bibnamefont
  {Carrasquilla}}, \bibinfo {author} {\bibfnamefont {Matthias}\ \bibnamefont
  {Troyer}}, \bibinfo {author} {\bibfnamefont {Roger}\ \bibnamefont {Melko}}, \
  and\ \bibinfo {author} {\bibfnamefont {Giuseppe}\ \bibnamefont {Carleo}},\
  }\bibfield  {title} {\enquote {\bibinfo {title} {Many-body quantum state
  tomography with neural networks},}\ }\href {https://arxiv.org/abs/1703.05334}
  {\bibfield  {journal} {\bibinfo  {journal} {Preprint: arXiv:1703.05334}\ }
  (\bibinfo {year} {2017})}\BibitemShut {NoStop}%
\bibitem [{\citenamefont {Deng}\ \emph {et~al.}(2016)\citenamefont {Deng},
  \citenamefont {Li},\ and\ \citenamefont {Sarma}}]{d_deng_16}%
  \BibitemOpen
  \bibfield  {author} {\bibinfo {author} {\bibfnamefont {Dong-Ling}\
  \bibnamefont {Deng}}, \bibinfo {author} {\bibfnamefont {Xiaopeng}\
  \bibnamefont {Li}}, \ and\ \bibinfo {author} {\bibfnamefont {S.~Das}\
  \bibnamefont {Sarma}},\ }\bibfield  {title} {\enquote {\bibinfo {title}
  {Exact machine learning topological states},}\ }\href
  {https://arxiv.org/pdf/1609.09060.pdf} {\  (\bibinfo {year} {2016})},\
  \Eprint {http://arxiv.org/abs/arXiv:1609.09060} {arXiv:1609.09060}
  \BibitemShut {NoStop}%
\bibitem [{\citenamefont {Mehta}\ and\ \citenamefont
  {Schwab}(2014)}]{p_mehta_14}%
  \BibitemOpen
  \bibfield  {author} {\bibinfo {author} {\bibfnamefont {Pankaj}\ \bibnamefont
  {Mehta}}\ and\ \bibinfo {author} {\bibfnamefont {David~J.}\ \bibnamefont
  {Schwab}},\ }\bibfield  {title} {\enquote {\bibinfo {title} {An exact mapping
  between the variational renormalization group and deep learning},}\ }\href
  {https://arxiv.org/abs/1410.3831} {\bibfield  {journal} {\bibinfo  {journal}
  {Preprint: arXiv:1410.3831}\ } (\bibinfo {year} {2014})}\BibitemShut
  {NoStop}%
\bibitem [{\citenamefont {Stoudenmire}\ and\ \citenamefont
  {Schwab}(2016)}]{m_stoudenmire_16}%
  \BibitemOpen
  \bibfield  {author} {\bibinfo {author} {\bibfnamefont {Edwin}\ \bibnamefont
  {Stoudenmire}}\ and\ \bibinfo {author} {\bibfnamefont {David~J}\ \bibnamefont
  {Schwab}},\ }\bibfield  {title} {\enquote {\bibinfo {title} {Supervised
  learning with tensor networks},}\ }in\ \href
  {http://papers.nips.cc/paper/6211-supervised-learning-with-tensor-networks.pdf}
  {\emph {\bibinfo {booktitle} {Advances in Neural Information Processing
  Systems 29}}},\ \bibinfo {editor} {edited by\ \bibinfo {editor}
  {\bibfnamefont {D.~D.}\ \bibnamefont {Lee}}, \bibinfo {editor} {\bibfnamefont
  {M.}~\bibnamefont {Sugiyama}}, \bibinfo {editor} {\bibfnamefont {U.~V.}\
  \bibnamefont {Luxburg}}, \bibinfo {editor} {\bibfnamefont {I.}~\bibnamefont
  {Guyon}}, \ and\ \bibinfo {editor} {\bibfnamefont {R.}~\bibnamefont
  {Garnett}}}\ (\bibinfo  {publisher} {Curran Associates, Inc.},\ \bibinfo
  {year} {2016})\ pp.\ \bibinfo {pages} {4799--4807}\BibitemShut {NoStop}%
\bibitem [{\citenamefont {Maaten}\ and\ \citenamefont
  {Hinton}(2008)}]{tSNE_Paper}%
  \BibitemOpen
  \bibfield  {author} {\bibinfo {author} {\bibfnamefont {L.v.d.}\ \bibnamefont
  {Maaten}}\ and\ \bibinfo {author} {\bibfnamefont {G.}~\bibnamefont
  {Hinton}},\ }\bibfield  {title} {\enquote {\bibinfo {title} {Visualizing data
  using t-sne},}\ }\href
  {http://www.jmlr.org/papers/volume9/vandermaaten08a/vandermaaten08a.pdf}
  {\bibfield  {journal} {\bibinfo  {journal} {Journal of Machine Learning
  Research}\ }\textbf {\bibinfo {volume} {9}},\ \bibinfo {pages} {2579Ñ2605}
  (\bibinfo {year} {2008})}\BibitemShut {NoStop}%
\bibitem [{tSN()}]{tSNE_Link}%
  \BibitemOpen
  \href@noop {} {\bibinfo  {journal} {A tutorial on how to use t-SNE
  effectively can be found at \url{https://distill.pub/2016/misread-tsne}}\
  }\BibitemShut {NoStop}%
\bibitem [{tSN(2017)}]{tSNE_Code}%
  \BibitemOpen
\bibfield  {journal} {  }\href@noop {} {\bibfield  {journal} {\bibinfo
  {journal} {Codes available at \url{https://lvdmaaten.github.io/tsne}}}}\BibitemShut {NoStop}%
\bibitem [{\citenamefont {Wang}(2016)}]{Wang2016}%
  \BibitemOpen
  \bibfield  {author} {\bibinfo {author} {\bibfnamefont {Lei}\ \bibnamefont
  {Wang}},\ }\bibfield  {title} {\enquote {\bibinfo {title} {Discovering phase
  transitions with unsupervised learning},}\ }\href {\doibase
  10.1103/PhysRevB.94.195105} {\bibfield  {journal} {\bibinfo  {journal} {Phys.
  Rev. B}\ }\textbf {\bibinfo {volume} {94}},\ \bibinfo {pages} {195105}
  (\bibinfo {year} {2016})}\BibitemShut {NoStop}%
\bibitem [{\citenamefont {Jolliffe}(2002)}]{pca}%
  \BibitemOpen
  \bibfield  {author} {\bibinfo {author} {\bibfnamefont {I.}~\bibnamefont
  {Jolliffe}},\ }\href@noop {} {\emph {\bibinfo {title} {Principal component
  analysis}}}\ (\bibinfo  {publisher} {John Wiley and Sons, Ltd},\ \bibinfo
  {year} {2002})\BibitemShut {NoStop}%
\bibitem [{\citenamefont {Wetzel}(2017)}]{s_wetzel_17}%
  \BibitemOpen
  \bibfield  {author} {\bibinfo {author} {\bibfnamefont {Sebastian~J.}\
  \bibnamefont {Wetzel}},\ }\bibfield  {title} {\enquote {\bibinfo {title}
  {Unsupervised learning of phase transitions: from principal component
  analysis to variational autoencoders},}\ }\href
  {https://arxiv.org/abs/1703.02435} {\bibfield  {journal} {\bibinfo  {journal}
  {Preprint: arXiv:1703.02435}\ } (\bibinfo {year} {2017})}\BibitemShut
  {NoStop}%
\bibitem [{\citenamefont {Hu}\ \emph {et~al.}(2017)\citenamefont {Hu},
  \citenamefont {Singh},\ and\ \citenamefont {Scalettar}}]{w_hu_17}%
  \BibitemOpen
  \bibfield  {author} {\bibinfo {author} {\bibfnamefont {Wenjian}\ \bibnamefont
  {Hu}}, \bibinfo {author} {\bibfnamefont {Rajiv R.~P.}\ \bibnamefont {Singh}},
  \ and\ \bibinfo {author} {\bibfnamefont {Richard~T.}\ \bibnamefont
  {Scalettar}},\ }\bibfield  {title} {\enquote {\bibinfo {title} {Discovering
  phases, phase transitions, and crossovers through unsupervised machine
  learning: A critical examination},}\ }\href {\doibase
  10.1103/PhysRevE.95.062122} {\bibfield  {journal} {\bibinfo  {journal} {Phys.
  Rev. E}\ }\textbf {\bibinfo {volume} {95}},\ \bibinfo {pages} {062122}
  (\bibinfo {year} {2017})}\BibitemShut {NoStop}%
\bibitem [{\citenamefont {Wang}\ and\ \citenamefont {Zhai}(2017)}]{c_wang_17}%
  \BibitemOpen
  \bibfield  {author} {\bibinfo {author} {\bibfnamefont {Ce}~\bibnamefont
  {Wang}}\ and\ \bibinfo {author} {\bibfnamefont {Hui}\ \bibnamefont {Zhai}},\
  }\bibfield  {title} {\enquote {\bibinfo {title} {Machine learning studies of
  frustrated classical spin models},}\ }\href
  {https://arxiv.org/abs/1706.07977} {\bibfield  {journal} {\bibinfo  {journal}
  {Preprint: arXiv:1706.07977}\ } (\bibinfo {year} {2017})}\BibitemShut
  {NoStop}%
\bibitem [{\citenamefont {Broecker}\ \emph {et~al.}(2017)\citenamefont
  {Broecker}, \citenamefont {Assaad},\ and\ \citenamefont
  {Trebst}}]{p_broecker_17}%
  \BibitemOpen
  \bibfield  {author} {\bibinfo {author} {\bibfnamefont {Peter}\ \bibnamefont
  {Broecker}}, \bibinfo {author} {\bibfnamefont {Fakher~F.}\ \bibnamefont
  {Assaad}}, \ and\ \bibinfo {author} {\bibfnamefont {Simon}\ \bibnamefont
  {Trebst}},\ }\bibfield  {title} {\enquote {\bibinfo {title} {Quantum phase
  recognition via unsupervised machine learning},}\ }\href
  {https://arxiv.org/pdf/1707.00663.pdf} {\bibfield  {journal} {\bibinfo
  {journal} {Preprint: arXiv:1707.00663}\ } (\bibinfo {year}
  {2017})}\BibitemShut {NoStop}%
\bibitem [{\citenamefont {Bourlard}\ and\ \citenamefont
  {Kamp}(1988)}]{h_bourlard_88}%
  \BibitemOpen
  \bibfield  {author} {\bibinfo {author} {\bibfnamefont {H.}~\bibnamefont
  {Bourlard}}\ and\ \bibinfo {author} {\bibfnamefont {Y.}~\bibnamefont
  {Kamp}},\ }\bibfield  {title} {\enquote {\bibinfo {title} {Auto-association
  by multilayer perceptrons and singular value decomposition},}\ }\href
  {\doibase 10.1007/BF00332918} {\bibfield  {journal} {\bibinfo  {journal}
  {Biological Cybernetics}\ }\textbf {\bibinfo {volume} {59}},\ \bibinfo
  {pages} {291--294} (\bibinfo {year} {1988})}\BibitemShut {NoStop}%
\bibitem [{\citenamefont {Hinton}\ and\ \citenamefont
  {Salakhutdinov}(2006)}]{g_hinton_06}%
  \BibitemOpen
  \bibfield  {author} {\bibinfo {author} {\bibfnamefont {G.~E.}\ \bibnamefont
  {Hinton}}\ and\ \bibinfo {author} {\bibfnamefont {R.~R.}\ \bibnamefont
  {Salakhutdinov}},\ }\bibfield  {title} {\enquote {\bibinfo {title} {Reducing
  the dimensionality of data with neural networks},}\ }\href {\doibase
  10.1126/science.1127647} {\bibfield  {journal} {\bibinfo  {journal}
  {Science}\ }\textbf {\bibinfo {volume} {313}},\ \bibinfo {pages} {504--507}
  (\bibinfo {year} {2006})}\BibitemShut {NoStop}%
\bibitem [{\citenamefont {Chollet}(2016)}]{f_chollet_16}%
  \BibitemOpen
  \bibfield  {author} {\bibinfo {author} {\bibfnamefont {F.}~\bibnamefont
  {Chollet}},\ }\bibfield  {title} {\enquote {\bibinfo {title} {Building
  autoencoders in keras},}\ }\href@noop {} {\bibfield  {journal} {\bibinfo
  {journal} {\url{https://blog.keras.io/building-autoencoders-in-keras.html}}\
  } (\bibinfo {year} {2016})}\BibitemShut {NoStop}%
\bibitem [{\citenamefont {Geurts}\ \emph {et~al.}(2006)\citenamefont {Geurts},
  \citenamefont {Ernst},\ and\ \citenamefont {Wehenkel}}]{p_geurts_06}%
  \BibitemOpen
  \bibfield  {author} {\bibinfo {author} {\bibfnamefont {Pierre}\ \bibnamefont
  {Geurts}}, \bibinfo {author} {\bibfnamefont {Damien}\ \bibnamefont {Ernst}},
  \ and\ \bibinfo {author} {\bibfnamefont {Louis}\ \bibnamefont {Wehenkel}},\
  }\bibfield  {title} {\enquote {\bibinfo {title} {Extremely randomized
  trees},}\ }\href {\doibase 10.1007/s10994-006-6226-1} {\bibfield  {journal}
  {\bibinfo  {journal} {Machine Learning}\ }\textbf {\bibinfo {volume} {63}},\
  \bibinfo {pages} {3--42} (\bibinfo {year} {2006})}\BibitemShut {NoStop}%
\bibitem [{\citenamefont {Moosmann}\ \emph {et~al.}(2006)\citenamefont
  {Moosmann}, \citenamefont {Triggs},\ and\ \citenamefont
  {Jurie}}]{f_moosmann_07}%
  \BibitemOpen
  \bibfield  {author} {\bibinfo {author} {\bibfnamefont {Frank}\ \bibnamefont
  {Moosmann}}, \bibinfo {author} {\bibfnamefont {Bill}\ \bibnamefont {Triggs}},
  \ and\ \bibinfo {author} {\bibfnamefont {Frederic}\ \bibnamefont {Jurie}},\
  }\bibfield  {title} {\enquote {\bibinfo {title} {Fast discriminative visual
  codebooks using randomized clustering forests},}\ }in\ \href
  {http://dl.acm.org/citation.cfm?id=2976456.2976580} {\emph {\bibinfo
  {booktitle} {Proceedings of the 19th International Conference on Neural
  Information Processing Systems}}},\ \bibinfo {series and number} {NIPS'06}\
  (\bibinfo  {publisher} {MIT Press},\ \bibinfo {address} {Cambridge, MA,
  USA},\ \bibinfo {year} {2006})\ pp.\ \bibinfo {pages} {985--992}\BibitemShut
  {NoStop}%
\bibitem [{ran()}]{randomtrees}%
  \BibitemOpen
  \href@noop {} {}\bibinfo {note} {Code availble through scikit-learn at
  \url{http://scikit-learn.org/stable/modules/generated/sklearn.ensemble.RandomTreesEmbedding.html}}\BibitemShut
  {NoStop}%
\bibitem [{\citenamefont {Hubbard}(1963)}]{j_hubbard_63}%
  \BibitemOpen
  \bibfield  {author} {\bibinfo {author} {\bibfnamefont {J.}~\bibnamefont
  {Hubbard}},\ }\bibfield  {title} {\enquote {\bibinfo {title} {Electron
  correlations in narrow energy bands},}\ }\href {\doibase
  10.1098/rspa.1963.0204} {\bibfield  {journal} {\bibinfo  {journal}
  {Proceedings of the Royal Society of London A: Mathematical, Physical and
  Engineering Sciences}\ }\textbf {\bibinfo {volume} {276}},\ \bibinfo {pages}
  {238--257} (\bibinfo {year} {1963})}\BibitemShut {NoStop}%
\bibitem [{\citenamefont {Loh}\ \emph {et~al.}(1990)\citenamefont {Loh},
  \citenamefont {Gubernatis}, \citenamefont {Scalettar}, \citenamefont {White},
  \citenamefont {Scalapino},\ and\ \citenamefont {Sugar}}]{e_loh_90}%
  \BibitemOpen
  \bibfield  {author} {\bibinfo {author} {\bibfnamefont {E.~Y.}\ \bibnamefont
  {Loh}}, \bibinfo {author} {\bibfnamefont {J.~E.}\ \bibnamefont {Gubernatis}},
  \bibinfo {author} {\bibfnamefont {R.~T.}\ \bibnamefont {Scalettar}}, \bibinfo
  {author} {\bibfnamefont {S.~R.}\ \bibnamefont {White}}, \bibinfo {author}
  {\bibfnamefont {D.~J.}\ \bibnamefont {Scalapino}}, \ and\ \bibinfo {author}
  {\bibfnamefont {R.~L.}\ \bibnamefont {Sugar}},\ }\bibfield  {title} {\enquote
  {\bibinfo {title} {Sign problem in the numerical simulation of many-electron
  systems},}\ }\href {\doibase 10.1103/PhysRevB.41.9301} {\bibfield  {journal}
  {\bibinfo  {journal} {Phys. Rev. B}\ }\textbf {\bibinfo {volume} {41}},\
  \bibinfo {pages} {9301--9307} (\bibinfo {year} {1990})}\BibitemShut {NoStop}%
\bibitem [{\citenamefont {Iglovikov}\ \emph {et~al.}(2015)\citenamefont
  {Iglovikov}, \citenamefont {Khatami},\ and\ \citenamefont
  {Scalettar}}]{v_iglovikov_15}%
  \BibitemOpen
  \bibfield  {author} {\bibinfo {author} {\bibfnamefont {V.~I.}\ \bibnamefont
  {Iglovikov}}, \bibinfo {author} {\bibfnamefont {E.}~\bibnamefont {Khatami}},
  \ and\ \bibinfo {author} {\bibfnamefont {R.~T.}\ \bibnamefont {Scalettar}},\
  }\bibfield  {title} {\enquote {\bibinfo {title} {Geometry dependence of the
  sign problem in quantum monte carlo simulations},}\ }\href {\doibase
  10.1103/PhysRevB.92.045110} {\bibfield  {journal} {\bibinfo  {journal} {Phys.
  Rev. B}\ }\textbf {\bibinfo {volume} {92}},\ \bibinfo {pages} {045110}
  (\bibinfo {year} {2015})}\BibitemShut {NoStop}%
\bibitem [{\citenamefont {Barber}\ \emph {et~al.}(1985)\citenamefont {Barber},
  \citenamefont {Pearson}, \citenamefont {Toussaint},\ and\ \citenamefont
  {Richardson}}]{3DIsing}%
  \BibitemOpen
  \bibfield  {author} {\bibinfo {author} {\bibfnamefont {Michael~N.}\
  \bibnamefont {Barber}}, \bibinfo {author} {\bibfnamefont {R.~B.}\
  \bibnamefont {Pearson}}, \bibinfo {author} {\bibfnamefont {Doug}\
  \bibnamefont {Toussaint}}, \ and\ \bibinfo {author} {\bibfnamefont {John~L.}\
  \bibnamefont {Richardson}},\ }\bibfield  {title} {\enquote {\bibinfo {title}
  {Finite-size scaling in the three-dimensional ising model},}\ }\href
  {\doibase 10.1103/PhysRevB.32.1720} {\bibfield  {journal} {\bibinfo
  {journal} {Phys. Rev. B}\ }\textbf {\bibinfo {volume} {32}},\ \bibinfo
  {pages} {1720--1730} (\bibinfo {year} {1985})}\BibitemShut {NoStop}%
\bibitem [{\citenamefont {Hastings}(1970)}]{w_hastings_70}%
  \BibitemOpen
  \bibfield  {author} {\bibinfo {author} {\bibfnamefont {W.~K.}\ \bibnamefont
  {Hastings}},\ }\bibfield  {title} {\enquote {\bibinfo {title} {Monte carlo
  sampling methods using markov chains and their applications},}\ }\href
  {\doibase 10.1093/biomet/57.1.97} {\bibfield  {journal} {\bibinfo  {journal}
  {Biometrika}\ }\textbf {\bibinfo {volume} {57}},\ \bibinfo {pages} {97--109}
  (\bibinfo {year} {1970})}\BibitemShut {NoStop}%
\bibitem [{\citenamefont {Scalettar}\ \emph {et~al.}(1989)\citenamefont
  {Scalettar}, \citenamefont {Scalapino}, \citenamefont {Sugar},\ and\
  \citenamefont {Toussaint}}]{r_scalettar_89}%
  \BibitemOpen
  \bibfield  {author} {\bibinfo {author} {\bibfnamefont {R.~T.}\ \bibnamefont
  {Scalettar}}, \bibinfo {author} {\bibfnamefont {D.~J.}\ \bibnamefont
  {Scalapino}}, \bibinfo {author} {\bibfnamefont {R.~L.}\ \bibnamefont
  {Sugar}}, \ and\ \bibinfo {author} {\bibfnamefont {D.}~\bibnamefont
  {Toussaint}},\ }\bibfield  {title} {\enquote {\bibinfo {title} {Phase diagram
  of the half-filled 3d hubbard model},}\ }\href {\doibase
  10.1103/PhysRevB.39.4711} {\bibfield  {journal} {\bibinfo  {journal} {Phys.
  Rev. B}\ }\textbf {\bibinfo {volume} {39}},\ \bibinfo {pages} {4711--4714}
  (\bibinfo {year} {1989})}\BibitemShut {NoStop}%
\bibitem [{\citenamefont {Staudt}\ \emph {et~al.}(2000)\citenamefont {Staudt},
  \citenamefont {Dzierzawa},\ and\ \citenamefont {Muramatsu}}]{r_staudt_00}%
  \BibitemOpen
  \bibfield  {author} {\bibinfo {author} {\bibfnamefont {R.}~\bibnamefont
  {Staudt}}, \bibinfo {author} {\bibfnamefont {M.}~\bibnamefont {Dzierzawa}}, \
  and\ \bibinfo {author} {\bibfnamefont {A.}~\bibnamefont {Muramatsu}},\
  }\bibfield  {title} {\enquote {\bibinfo {title} {Phase diagram of the
  three-dimensional hubbard model at half filling},}\ }\href {\doibase
  10.1007/s100510070120} {\bibfield  {journal} {\bibinfo  {journal} {The
  European Physical Journal B - Condensed Matter and Complex Systems}\ }\textbf
  {\bibinfo {volume} {17}},\ \bibinfo {pages} {411--415} (\bibinfo {year}
  {2000})}\BibitemShut {NoStop}%
\bibitem [{\citenamefont {Kent}\ \emph {et~al.}(2005)\citenamefont {Kent},
  \citenamefont {Jarrell}, \citenamefont {Maier},\ and\ \citenamefont
  {Pruschke}}]{p_kent_05}%
  \BibitemOpen
  \bibfield  {author} {\bibinfo {author} {\bibfnamefont {P.~R.~C.}\
  \bibnamefont {Kent}}, \bibinfo {author} {\bibfnamefont {M.}~\bibnamefont
  {Jarrell}}, \bibinfo {author} {\bibfnamefont {T.~A.}\ \bibnamefont {Maier}},
  \ and\ \bibinfo {author} {\bibfnamefont {Th.}\ \bibnamefont {Pruschke}},\
  }\bibfield  {title} {\enquote {\bibinfo {title} {Efficient calculation of the
  antiferromagnetic phase diagram of the three-dimensional hubbard model},}\
  }\href {\doibase 10.1103/PhysRevB.72.060411} {\bibfield  {journal} {\bibinfo
  {journal} {Phys. Rev. B}\ }\textbf {\bibinfo {volume} {72}},\ \bibinfo
  {pages} {060411} (\bibinfo {year} {2005})}\BibitemShut {NoStop}%
\bibitem [{\citenamefont {Paiva}\ \emph {et~al.}(2011)\citenamefont {Paiva},
  \citenamefont {Loh}, \citenamefont {Randeria}, \citenamefont {Scalettar},\
  and\ \citenamefont {Trivedi}}]{t_paiva_11}%
  \BibitemOpen
  \bibfield  {author} {\bibinfo {author} {\bibfnamefont {Thereza}\ \bibnamefont
  {Paiva}}, \bibinfo {author} {\bibfnamefont {Yen~Lee}\ \bibnamefont {Loh}},
  \bibinfo {author} {\bibfnamefont {Mohit}\ \bibnamefont {Randeria}}, \bibinfo
  {author} {\bibfnamefont {Richard~T.}\ \bibnamefont {Scalettar}}, \ and\
  \bibinfo {author} {\bibfnamefont {Nandini}\ \bibnamefont {Trivedi}},\
  }\bibfield  {title} {\enquote {\bibinfo {title} {Fermions in 3d optical
  lattices: Cooling protocol to obtain antiferromagnetism},}\ }\href {\doibase
  10.1103/PhysRevLett.107.086401} {\bibfield  {journal} {\bibinfo  {journal}
  {Phys. Rev. Lett.}\ }\textbf {\bibinfo {volume} {107}},\ \bibinfo {pages}
  {086401} (\bibinfo {year} {2011})}\BibitemShut {NoStop}%
\bibitem [{\citenamefont {Kozik}\ \emph {et~al.}(2013)\citenamefont {Kozik},
  \citenamefont {Burovski}, \citenamefont {Scarola},\ and\ \citenamefont
  {Troyer}}]{e_kozik_13}%
  \BibitemOpen
  \bibfield  {author} {\bibinfo {author} {\bibfnamefont {E.}~\bibnamefont
  {Kozik}}, \bibinfo {author} {\bibfnamefont {E.}~\bibnamefont {Burovski}},
  \bibinfo {author} {\bibfnamefont {V.~W.}\ \bibnamefont {Scarola}}, \ and\
  \bibinfo {author} {\bibfnamefont {M.}~\bibnamefont {Troyer}},\ }\bibfield
  {title} {\enquote {\bibinfo {title} {N\'eel temperature and thermodynamics of
  the half-filled three-dimensional hubbard model by diagrammatic determinant
  monte carlo},}\ }\href {\doibase 10.1103/PhysRevB.87.205102} {\bibfield
  {journal} {\bibinfo  {journal} {Phys. Rev. B}\ }\textbf {\bibinfo {volume}
  {87}},\ \bibinfo {pages} {205102} (\bibinfo {year} {2013})}\BibitemShut
  {NoStop}%
\bibitem [{\citenamefont {Hirschmeier}\ \emph {et~al.}(2015)\citenamefont
  {Hirschmeier}, \citenamefont {Hafermann}, \citenamefont {Gull}, \citenamefont
  {Lichtenstein},\ and\ \citenamefont {Antipov}}]{d_hirschmeier_15}%
  \BibitemOpen
  \bibfield  {author} {\bibinfo {author} {\bibfnamefont {Daniel}\ \bibnamefont
  {Hirschmeier}}, \bibinfo {author} {\bibfnamefont {Hartmut}\ \bibnamefont
  {Hafermann}}, \bibinfo {author} {\bibfnamefont {Emanuel}\ \bibnamefont
  {Gull}}, \bibinfo {author} {\bibfnamefont {Alexander~I.}\ \bibnamefont
  {Lichtenstein}}, \ and\ \bibinfo {author} {\bibfnamefont {Andrey~E.}\
  \bibnamefont {Antipov}},\ }\bibfield  {title} {\enquote {\bibinfo {title}
  {Mechanisms of finite-temperature magnetism in the three-dimensional hubbard
  model},}\ }\href {\doibase 10.1103/PhysRevB.92.144409} {\bibfield  {journal}
  {\bibinfo  {journal} {Phys. Rev. B}\ }\textbf {\bibinfo {volume} {92}},\
  \bibinfo {pages} {144409} (\bibinfo {year} {2015})}\BibitemShut {NoStop}%
\bibitem [{\citenamefont {Khatami}(2016)}]{e_khatami_16}%
  \BibitemOpen
  \bibfield  {author} {\bibinfo {author} {\bibfnamefont {Ehsan}\ \bibnamefont
  {Khatami}},\ }\bibfield  {title} {\enquote {\bibinfo {title}
  {Three-dimensional hubbard model in the thermodynamic limit},}\ }\href
  {\doibase 10.1103/PhysRevB.94.125114} {\bibfield  {journal} {\bibinfo
  {journal} {Phys. Rev. B}\ }\textbf {\bibinfo {volume} {94}},\ \bibinfo
  {pages} {125114} (\bibinfo {year} {2016})}\BibitemShut {NoStop}%
\bibitem [{\citenamefont {Sandvik}(1998)}]{a_sandvik_98}%
  \BibitemOpen
  \bibfield  {author} {\bibinfo {author} {\bibfnamefont {Anders~W.}\
  \bibnamefont {Sandvik}},\ }\bibfield  {title} {\enquote {\bibinfo {title}
  {Critical temperature and the transition from quantum to classical order
  parameter fluctuations in the three-dimensional heisenberg
  antiferromagnet},}\ }\href {\doibase 10.1103/PhysRevLett.80.5196} {\bibfield
  {journal} {\bibinfo  {journal} {Phys. Rev. Lett.}\ }\textbf {\bibinfo
  {volume} {80}},\ \bibinfo {pages} {5196--5199} (\bibinfo {year}
  {1998})}\BibitemShut {NoStop}%
\bibitem [{\citenamefont {Hart}\ \emph {et~al.}(2015)\citenamefont {Hart},
  \citenamefont {Duarte}, \citenamefont {Yang}, \citenamefont {Liu},
  \citenamefont {Paiva}, \citenamefont {Khatami}, \citenamefont {Scalettar},
  \citenamefont {Trivedi}, \citenamefont {Huse},\ and\ \citenamefont
  {Hulet}}]{r_hart_15}%
  \BibitemOpen
  \bibfield  {author} {\bibinfo {author} {\bibfnamefont {R.~A.}\ \bibnamefont
  {Hart}}, \bibinfo {author} {\bibfnamefont {P.~M.}\ \bibnamefont {Duarte}},
  \bibinfo {author} {\bibfnamefont {T.~L.}\ \bibnamefont {Yang}}, \bibinfo
  {author} {\bibfnamefont {X.}~\bibnamefont {Liu}}, \bibinfo {author}
  {\bibfnamefont {T.}~\bibnamefont {Paiva}}, \bibinfo {author} {\bibfnamefont
  {E.}~\bibnamefont {Khatami}}, \bibinfo {author} {\bibfnamefont {R.~T.}\
  \bibnamefont {Scalettar}}, \bibinfo {author} {\bibfnamefont {N.}~\bibnamefont
  {Trivedi}}, \bibinfo {author} {\bibfnamefont {D.~A.}\ \bibnamefont {Huse}}, \
  and\ \bibinfo {author} {\bibfnamefont {R.~G.}\ \bibnamefont {Hulet}},\
  }\bibfield  {title} {\enquote {\bibinfo {title} {Observation of
  antiferromagnetic correlations in the hubbard model with ultracold atoms},}\
  }\href {http://dx.doi.org/10.1038/nature14223} {\bibfield  {journal}
  {\bibinfo  {journal} {Nature}\ }\textbf {\bibinfo {volume} {519}},\ \bibinfo
  {pages} {211--214} (\bibinfo {year} {2015})}\BibitemShut {NoStop}%
\bibitem [{\citenamefont {Blankenbecler}\ \emph {et~al.}(1981)\citenamefont
  {Blankenbecler}, \citenamefont {Scalapino},\ and\ \citenamefont
  {Sugar}}]{r_blankenbecler_81}%
  \BibitemOpen
  \bibfield  {author} {\bibinfo {author} {\bibfnamefont {R.}~\bibnamefont
  {Blankenbecler}}, \bibinfo {author} {\bibfnamefont {D.~J.}\ \bibnamefont
  {Scalapino}}, \ and\ \bibinfo {author} {\bibfnamefont {R.~L.}\ \bibnamefont
  {Sugar}},\ }\bibfield  {title} {\enquote {\bibinfo {title} {Monte carlo
  calculations of coupled boson-fermion systems. i},}\ }\href {\doibase
  10.1103/PhysRevD.24.2278} {\bibfield  {journal} {\bibinfo  {journal} {Phys.
  Rev. D}\ }\textbf {\bibinfo {volume} {24}},\ \bibinfo {pages} {2278--2286}
  (\bibinfo {year} {1981})}\BibitemShut {NoStop}%
\bibitem [{que()}]{quest}%
  \BibitemOpen
  \href@noop {} {\bibinfo  {journal} {Codes available at
  \url{http://quest.ucdavis.edu/} and
  \url{https://code.google.com/p/quest-qmc}}\ }\BibitemShut {NoStop}%
\bibitem [{\citenamefont {Paiva}\ \emph {et~al.}(2010)\citenamefont {Paiva},
  \citenamefont {Scalettar}, \citenamefont {Randeria},\ and\ \citenamefont
  {Trivedi}}]{t_paiva_10}%
  \BibitemOpen
\bibfield  {journal} {  }\bibfield  {author} {\bibinfo {author} {\bibfnamefont
  {Thereza}\ \bibnamefont {Paiva}}, \bibinfo {author} {\bibfnamefont {Richard}\
  \bibnamefont {Scalettar}}, \bibinfo {author} {\bibfnamefont {Mohit}\
  \bibnamefont {Randeria}}, \ and\ \bibinfo {author} {\bibfnamefont {Nandini}\
  \bibnamefont {Trivedi}},\ }\bibfield  {title} {\enquote {\bibinfo {title}
  {Fermions in 2d optical lattices: Temperature and entropy scales for
  observing antiferromagnetism and superfluidity},}\ }\href {\doibase
  10.1103/PhysRevLett.104.066406} {\bibfield  {journal} {\bibinfo  {journal}
  {Phys. Rev. Lett.}\ }\textbf {\bibinfo {volume} {104}},\ \bibinfo {pages}
  {066406} (\bibinfo {year} {2010})}\BibitemShut {NoStop}%
\bibitem [{\citenamefont {Khatami}\ and\ \citenamefont
  {M.{\color{white}~}Rigol}(2011)}]{E_khatami_11b}%
  \BibitemOpen
  \bibfield  {author} {\bibinfo {author} {\bibfnamefont {Ehsan}\ \bibnamefont
  {Khatami}}\ and\ \bibinfo {author} {\bibnamefont {M.{\color{white}~}Rigol}},\
  }\bibfield  {title} {\enquote {\bibinfo {title} {Thermodynamics of strongly
  interacting fermions in two-dimensional optical lattices},}\ }\href {\doibase
  10.1103/PhysRevA.84.053611} {\bibfield  {journal} {\bibinfo  {journal} {Phys.
  Rev. A}\ }\textbf {\bibinfo {volume} {84}},\ \bibinfo {pages} {053611}
  (\bibinfo {year} {2011})}\BibitemShut {NoStop}%
\bibitem [{\citenamefont {Wikipedia}(2017)}]{ANN}%
  \BibitemOpen
  \bibfield  {author} {\bibinfo {author} {\bibnamefont {Wikipedia}},\ }\href
  {https://en.wikipedia.org/w/index.php?title=Artificial_neural_network&oldid=794334514}
  {\enquote {\bibinfo {title} {Artificial neural network --- wikipedia{,} the
  free encyclopedia},}\ } (\bibinfo {year} {2017}),\ \bibinfo {note} {[Online;
  accessed 9-August-2017]}\BibitemShut {NoStop}%
\bibitem [{\citenamefont {Baldi}\ and\ \citenamefont
  {Hornik}(1989)}]{p_baldi_89}%
  \BibitemOpen
  \bibfield  {author} {\bibinfo {author} {\bibfnamefont {Pierre}\ \bibnamefont
  {Baldi}}\ and\ \bibinfo {author} {\bibfnamefont {Kurt}\ \bibnamefont
  {Hornik}},\ }\bibfield  {title} {\enquote {\bibinfo {title} {Neural networks
  and principal component analysis: Learning from examples without local
  minima},}\ }\href {\doibase http://dx.doi.org/10.1016/0893-6080(89)90014-2}
  {\bibfield  {journal} {\bibinfo  {journal} {Neural Networks}\ }\textbf
  {\bibinfo {volume} {2}},\ \bibinfo {pages} {53 -- 58} (\bibinfo {year}
  {1989})}\BibitemShut {NoStop}%
\bibitem [{\citenamefont {Scalettar}()}]{ScalettarBeijing}%
  \BibitemOpen
  \bibfield  {author} {\bibinfo {author} {\bibfnamefont {Richard}\ \bibnamefont
  {Scalettar}},\ }\href@noop {} {}\bibinfo {note} {Talk at the Machine Learning
  and Many-body Physics conference at KTIS, Beijing, China.
  \url{http://kits.ucas.ac.cn/images/articles/2017/Machine\_Learning/conference/10RichardScalettar.pdf}}\BibitemShut
  {NoStop}%
\bibitem [{\citenamefont {Hinton}\ and\ \citenamefont
  {Roweis}(2003)}]{g_hinton_03}%
  \BibitemOpen
  \bibfield  {author} {\bibinfo {author} {\bibfnamefont {Geoffrey~E}\
  \bibnamefont {Hinton}}\ and\ \bibinfo {author} {\bibfnamefont {Sam~T}\
  \bibnamefont {Roweis}},\ }\bibfield  {title} {\enquote {\bibinfo {title}
  {Stochastic neighbor embedding},}\ }in\ \href@noop {} {\emph {\bibinfo
  {booktitle} {Advances in neural information processing systems}}}\ (\bibinfo
  {year} {2003})\ pp.\ \bibinfo {pages} {857--864}\BibitemShut {NoStop}%
\bibitem [{\citenamefont {Yang}\ \emph {et~al.}(2013)\citenamefont {Yang},
  \citenamefont {Peltonen},\ and\ \citenamefont {Kaski}}]{z_yang_13}%
  \BibitemOpen
  \bibfield  {author} {\bibinfo {author} {\bibfnamefont {Zhirong}\ \bibnamefont
  {Yang}}, \bibinfo {author} {\bibfnamefont {Jaakko}\ \bibnamefont {Peltonen}},
  \ and\ \bibinfo {author} {\bibfnamefont {Samuel}\ \bibnamefont {Kaski}},\
  }\bibfield  {title} {\enquote {\bibinfo {title} {Scalable optimization of
  neighbor embedding for visualization},}\ }in\ \href
  {http://proceedings.mlr.press/v28/yang13b.html} {\emph {\bibinfo {booktitle}
  {Proceedings of the 30th International Conference on Machine Learning}}},\
  \bibinfo {series} {Proceedings of Machine Learning Research}, Vol.~\bibinfo
  {volume} {28},\ \bibinfo {editor} {edited by\ \bibinfo {editor}
  {\bibfnamefont {Sanjoy}\ \bibnamefont {Dasgupta}}\ and\ \bibinfo {editor}
  {\bibfnamefont {David}\ \bibnamefont {McAllester}}}\ (\bibinfo  {publisher}
  {PMLR},\ \bibinfo {address} {Atlanta, Georgia, USA},\ \bibinfo {year}
  {2013})\ pp.\ \bibinfo {pages} {127--135}\BibitemShut {NoStop}%
\bibitem [{AFM()}]{AFMSF}%
  \BibitemOpen
  \href@noop {} {}\bibinfo {note} {The structure factor is defined as $S^z_{\bf
  Q}=1/N\sum_{ij}e^{i{\bf Q}\cdot{\bf R}_{ij}}S^z_iS^z_j$, where $N$ is the
  system size, $S^z_i=(n_{i\uparrow}-n_{i\downarrow})/2$ is the $z$ component
  of spin at site $i$, ${\bf R}_{ij}$ is the vector connecting sites $i$ and
  $i$, and ${\bf Q}$ is the wavevector. The latter is 0 for the ferromagnetic
  structure factor and $(\pi,\pi,\pi)$ or $(\pi,\pi)$ for the AFM structure
  factor in 3D or 2D, respectively. In the absence of any SU(2) symmetry
  breaking terms in the Hamiltonian, this structure factor is equal to
  $S^{xy}_{\bf Q}=1/2N\sum_{ij}e^{i{\bf Q}\cdot{\bf R}_{ij}}S^+_iS^-_j$, where
  $S^{\pm}_i$ are the spin raising/lowering operators.}\BibitemShut {Stop}%
\end{thebibliography}

%

\end{document}